\documentclass[floatfix,prb,twocolumn]{revtex4}
\usepackage{longtable}
\usepackage{graphicx}
\usepackage{amsmath}
\usepackage{amssymb}
\usepackage{multirow}
\usepackage{bm}

\DeclareMathOperator{\sgn}{sgn}
\DeclareMathOperator{\Cl}{Cl_{2}}

\DeclareMathOperator{\Real}{Re}
\DeclareMathOperator{\Li}{Li_{2}}
\DeclareMathOperator{\atan}{arctan}

\begin{document}

\title{Elastic Theory of Defects in Toroidal Crystals}

\author{Luca Giomi}
\email{lgiomi@physics.syr.edu}

\author{Mark J. Bowick}
\email{bowick@physics.syr.edu}

\affiliation{Department of Physics,
Syracuse University,
Syracuse New York,
13244-1130}

\begin{abstract}
We report a comprehensive analysis of the ground state properties of 
axisymmetric toroidal crystals based on the elastic theory of defects
on curved substrates. The ground state is analyzed as a function of 
the aspect ratio of the torus, which provides a non-local measure of 
the underlying Gaussian curvature, and the ratio of the defect 
core-energy to the Young modulus. Several structural features are discussed,
including a spectacular example of curvature-driven amorphization in
the limit of the aspect ratio approaching one. The outcome of the elastic 
theory is then compared with the results of a numerical study of a system 
of point-like particles constrained on the surface of a torus and
interacting via a short range potential. 
\end{abstract}

\maketitle

\section{\label{sec:1}Introduction}

The study of the spontaneous organization of microscopic objects
such as colloids, amphiphiles or protein clusters into mesoscale 
structures (mesoatoms) is an active research area that offers challenges 
on many fronts \cite{Glotzer:04,WB:02}. One would like to create a rich
warehouse of raw materials from which to engineer mesomolecules or
bulk materials with novel mechanical, optical or electronic behavior.

A fertile source of potential mesoatoms is provided by the self-assembly 
of micron or smaller scale colloidal particles on {\em curved}
interfaces. The interplay between spatial curvature and condensed
matter order in these systems has proven to be very rich. Spatial 
curvature leads to novel defect arrays in the ground state that 
have widespread implications for the fundamental and applied physics 
of curved phases of matter with crystalline, hexatic or nematic order.

In a recent paper Kim \emph{et al} reported the formation of
toroidal micelles from the self-assembly of dumbbell-shaped
amphiphilic molecules \cite{KimEtAl:2006}. Molecular dumbbells
dissolved in a selective solvent self-assemble in an aggregate
structure due to their amphiphilic character. This process has been
observed to yield coexisting spherical and open-ended cylindrical
micelles. These structures change slowly over the course of a
week to toroidal micelles which thus appear more stable. Toroidal
geometries also occur in microbiology in the viral capsid of the
coronavirus \emph{torovirus} \cite{SnijderHorzinek}. The torovirus
is an RNA viral package of maximal diameter between $120$ and $140$
nm and is surrounded, as other coronaviridae, by a double
wreath/ring of cladding proteins.

Carbon nanotori form another fascinating and technologically
promising class of toroidal crystals \cite{LiuEtAl:1997} with
remarkable magnetic and electronic properties. The interplay between
the ballistic motion of the $\pi$ electrons and the geometry of the
embedding torus leads to a rich variety of quantum mechanical
properties including Pauli paramagnetism \cite{MeunierEtAl:1998} and 
Aharonov-Bohm oscillations in the magnetization \cite{LiuChenDing:2008}.
Ring closure of carbon nanotubes by chemical methods \cite{SanoEtAl:2001}
suggest that nanotubes may be more flexible than at first thought
and provides another technique of constructing carbon tori.

A unified theoretical framework to describe the structure of
toroidal crystals is provided by the elastic theory of defects in a
curved background \cite{PerezGarridoEtAl:1997,BowickNelsonTravesset:2000,
VitelliLucksNelson:2006,GiomiBowick:2007}. This formalism has 
the advantage of far fewer degrees of freedom than a direct 
treatment of the microscopic interactions and allows one to 
explore the origin of the emergent symmetry observed in toroidal 
crystals as the result of the interplay between defects and 
geometry. The latter is one of the fundamental hallmarks 
of two-dimensional non-Euclidean crystals and leads to
universal features observed in systems as different as viral 
capsids and carbon macromolecules.

In this paper we provide a detailed analysis of the structural 
properties of toroidal crystals. We show that the ground state 
has excess $5-$fold disclination defects on the exterior of the 
torus and $7-$fold defects on the interior. The precise number of 
excess disclinations, as well as their arrangement, is primarily 
controlled by the aspect ratio of the torus. Since defective 
regions are physically distinguished locations they are natural 
places for biological activity and chemical functionalization. 
A thorough understanding of the surface structure of crystalline 
assemblages could represent a significant step towards a 
first-principles design of entire libraries of nano and mesoscale 
components with precisely determined valence.  Such mesoatoms 
could serve in turn as the building blocks for mesomolecules or 
bulk materials via spontaneous self-assembly or controlled 
fabrication.

The paper is organized as follows. In Sec. \ref{sec:2} we 
review the elastic theory of defects in two-dimensional curved 
geometries and derive the ground state energy of a toroidal 
crystal. In Sec. \ref{sec:3} we summarize some fundamentals of 
the geometry of triangulated tori and discuss how the 
intrinsically discrete problem of crystallography can be 
reconciled with the result of the continuous elastic theory 
presented in the previous section. Sec. \ref{sec:4} is devoted 
to the analysis of the crystalline structure arising from the 
solution of the elastic problem. In Sec. \ref{sec:5} we discuss 
the results of numerical minimization of the potential energy of 
a system of classical particles interacting via a short-range 
potential on the surface of a torus in the light of the results 
of Sec. \ref{sec:4}. Finally, in Sec. \ref{sec:6} we specialize 
our analysis to the case of a ``fat torus'' of aspect ratio one 
and we show how the curvature singularity at the center of the 
torus is responsible for a remarkable curvature-driven transition 
to a disordered state. Some of the results presented here have
been already announced in Ref. \onlinecite{GiomiBowick:2008}.

\section{\label{sec:2}Defects and Geometry}

A two-dimensional torus of revolution $T^2$ is described in 
parametric form by:
\begin{equation}
\left\{
\begin{array}{l}
x = (R_{1}+R_{2}\cos\psi)\cos\phi\\
y = (R_{1}+R_{2}\cos\psi)\sin\phi\\
z = R_{2}\sin\psi
\end{array}
\right.\,,
\end{equation}
where $R_{1}>R_{2}$ are the two radii of the torus. The metric 
tensor $g_{ij}$ (with determinant $g$) and the Gaussian curvature 
$K$ can be written respectively as:
\begin{gather}
g_{ij} =
\left(
\begin{array}{cc}
R_{2}^{2} & 0 \\
0        & (R_{1}+R_{2}\cos\psi)^{2}
\end{array}
\right)\,,\\[7pt]
K = \frac{\cos\psi}{R_{2}(R_{1}+R_{2}\cos\psi)}\,.
\end{gather}
Within the elastic theory of defects on curved surfaces the original 
interacting particle problem is mapped to a system of interacting 
disclination defects in a continuum elastic curved background. 
Disclinations are characterized by their topological or disclination 
charge, $q_{i}$, representing the departure of a vertex from a the $6-$fold
coordination of a perfect triangular lattice. Thus $q_{i}=6-c_{i}$, where $c_{i}$ 
is the coordination number of the $i$th vertex. A classic theorem of Euler requires 
the total disclination charge of any triangulation of a two-dimensional Riemannian 
manifold $M$ to equal $6\chi_{M}$, where $\chi_{M}$ is the Euler 
characteristic of $M$. In the case of the torus $\chi_{M}=0$, and thus 
disclinations must appear in pairs of opposite disclination charge (i.e. 
$5-$fold and $7-$fold vertices with $q_{i}=1$ and $-1$ respectively) in 
order to ensure disclination charge neutrality. 
\begin{figure}
\centering
\includegraphics[width=0.8\columnwidth]{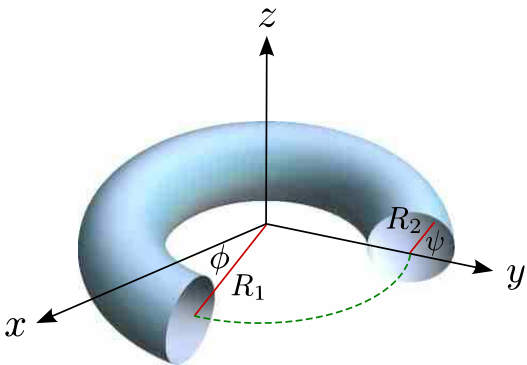}
\caption{\label{fig:torus-section}(Color online) The standard parametrization of 
a circular torus of radii $R_{1}$ and $R_{2}$.}
\end{figure}

The total free energy of a toroidal crystal with $N$ disclinations can be 
expressed as
\begin{equation}\label{eq:free_energy}
F = \frac{1}{2Y}\int d^{2}x\,\Gamma^{2}(\bm{x})+\epsilon_{c}\sum_{i=1}^{N}q_{i}^{2}+F_{0}\,,
\end{equation}
where $Y$ is the 2D Young modulus and $\Gamma(\bm{x})$ is the
solution of the following Poisson problem with periodic boundary
conditions:
\begin{equation}\label{eq:poisson}
\Delta_{g}\Gamma(\bm{x})=Y\rho(\bm{x})\,,
\end{equation}
where $\Delta_{g}$ is the Laplace-Beltrami operator 
\begin{equation}
\Delta_{g} 
= g^{ij}\nabla_{i}\nabla_{j} 
= \frac{1}{\sqrt{g}}\,\partial_{i}\sqrt{g}\,g^{ij}\partial_{j}
\end{equation}
and $\rho(\bm{x})$ is the total topological charge density
\begin{equation}\label{eq:charge_density}
\rho(\bm{x})=\frac{\pi}{3}\sum_{k=1}^{N}q_{k}\delta(\bm{x},\bm{x}_{k})-K(\bm{x})
\end{equation}
of $N$ disclinations located at the sites $\bm{x}_{k}$ together with the
screening contribution due to the Gaussian curvature $K(\bm{x})$ of the 
embedding manifold. 

The first term in Eq.~\eqref{eq:free_energy} represents the long-range 
elastic distortion due to defects and curvature. The second term in Eq. 
\eqref{eq:free_energy} is the defect core-energy representing the energy 
required to create a single disclination defect. This quantity is related 
to the short-distance cut-off of the elastic theory and is proportional to 
the square of the topological charge times a constant 
$\epsilon_{c}$~\cite{NelsonHalperin:1979}. Finally $F_{0}$ is the free 
energy of a flat defect-free monolayer. The Gauss-Bonnet theorem~\cite{DoCarmo} 
requires the total topological charge to be zero on the torus:
\begin{equation}\label{eq:gauss_bonnet}
\sum_{k=1}^{N}q_{k} = \int d^{2}x\, K(\bm{x}) = 0
\end{equation}
The function $\Gamma(\bm{x})$ can be expressed in the Green form:
\begin{equation}\label{eq:gamma}
\frac{\Gamma(\bm{x})}{Y} 
= \int d^{2}y\,G_{L}(\bm{x},\bm{y})\rho(\bm{y})\,,
\end{equation}
where $G_{L}(\bm{x},\bm{y})$, the Laplacian Green function on the torus, 
satisfies the equation:
\begin{equation}\label{eq:green_equation}
\Delta_{g}G_{L}(\bm{x},\bm{y})=\delta_{g}(\bm{x},\bm{y})-A^{-1}\,,
\end{equation}
with $\delta_{g}(\bm{x},\bm{y})$ the Dirac delta function on the torus and 
$A=4\pi^{2}R_{1}R_{2}$ is the surface area. It is easy to prove that the 
solution of the traditional Green-Laplace equation doesn't exist on closed 
manifolds like the torus, from which the extra term $A^{-1}$ appearing in Eq. 
\eqref{eq:green_equation}. $A^{-1}$ is the eigenfunction of the Laplacian 
associated with the null eigenvalue (zero mode) and is necessary
because a pure isolated source (giving rise to the delta-function 
on the right-hand side) has no place to terminate the field on the
closed torus.

As usual the calculation of the Green function can be remarkably simplified 
by conformally mapping the torus to a domain of the Euclidean plane via a 
suitable system of isothermal coordinates. Intuitively the torus is 
conformally equivalent to a rectangular domain described by a system of 
Cartesian coordinates. To make this explicit, one can equate the 
metric of the torus in the coordinates $(\psi,\phi)$ to a conformally 
Euclidean metric in the coordinates $(\xi,\eta)$:
\[
ds^{2}
= R_{2}^{2}d\psi^{2}+(R_{1}+R_{2}\cos\psi)^{2}d\phi^{2}
= w\,(d\xi^{2}+d\eta^{2})\,,
\]
where $w$ is a positive conformal factor. Taking $\eta=\phi$ and 
$w=(R_{1}+R_{2}\cos\psi)^{2}$, the coordinate $\xi$ is determined by 
the differential equation:
\begin{equation}\label{eq:conformal_ode}
\frac{d\xi}{d\psi}
= \pm \frac{1}{r+\cos\psi}\,,
\end{equation}
where $r=R_{1}/R_{2}$, the aspect ratio of the torus, may be taken greater
or equal to one without loss of generality. Choosing the plus sign and 
integrating both sides of Eq.~\eqref{eq:conformal_ode} we find:
\begin{equation}\label{eq:xi}
\xi = \int_{0}^{\psi}\frac{d\psi'}{r+\cos\psi'}\,.
\end{equation}
Taking $\psi\in[-\pi,\pi]$, the integral \eqref{eq:xi} yields:
\[
\xi = \kappa\atan\left(\omega\tan\frac{\psi}{2}\right)\,,
\]
where
\begin{equation}\label{eq:kappa-omega}
\kappa = \frac{2}{\sqrt{r^{2}-1}}\,,
\qquad\qquad
\omega = \sqrt{\frac{r-1}{r+1}}\,.
\end{equation}
In the transformed coordinate system $(\xi,\eta)$ the modified Green-Laplace 
equation reads:
\begin{equation}\label{eq:green3}
\Delta G_{L}(\bm{x},\bm{y}) = \delta(\bm{x},\bm{y})-\frac{w}{A}\,,
\end{equation}
where $\Delta$ and $\delta$ are now the Euclidean Laplacian and delta function.
The function $G_{L}(\bm{x},\bm{y})$ can be expressed in the form:
\[
G_{L}(\bm{x},\bm{y})
= G_{0}(\bm{x},\bm{y})
-\langle G_{0}(\bm{x},\cdot\,)\rangle 
-\langle G_{0}(\cdot\,,\bm{y})\rangle 
+\langle G_{0}(\cdot\,,\cdot\,) \rangle\,,
\]
where $G_{0}(\bm{x},\bm{y})$ is the Laplacian Green function on a periodic 
rectangle and the angular brackets stand for the normalized integral of the 
function $G_{0}(\bm{x},\bm{y})$ with respect to the dotted variable:
\begin{equation}
\langle G_{0}(\bm{x},\cdot\,)\rangle = \int \frac{d^{2}y}{A}\,G_{0}(\bm{x},\bm{y})\,. 
\end{equation}
Analogously the function $\langle G_{0}(\cdot\,,\cdot\,) \rangle$ is given by
\[
\langle G_{0}(\cdot\,,\cdot\,) \rangle = \int \frac{d^{2}x\,d^{2}y}{A^{2}}\,G_{0}(\bm{x},\bm{y}) 
\]
and ensures the neutrality property:
\begin{equation}\label{eq:green_neutrality}
\int d^{2}x\,G_{L}(\bm{x},\bm{y}) = \int d^{2}y\,G_{L}(\bm{x},\bm{y}) = 0\,.
\end{equation}

Using standard analysis the Green function $G_{0}(\bm{x},\bm{y})$ at the 
points $\bm{x}=(\xi,\eta)$ and $\bm{y}=(\xi',\eta')$ of a periodic rectangle 
of edges $p_{1}$ and $p_{2}$ can be calculated in the form:
\begin{multline}\label{eq:periodic_green_function}
G_{0}(\bm{x},\bm{y})
= \frac{\log 2}{6\pi}
-\frac{1}{2\,p_{1}p_{2}}|\eta-\eta'|^{2}\\[7pt]
+\frac{1}{2\pi}\log\left|\frac{\vartheta_{1}(\frac{z-z'}{p_{1}/\pi}|\frac{ip_{2}}{p_{1}})}
 {\vartheta_{1}'^{\frac{1}{3}}(0|\frac{ip_{2}}{p_{1}})}\right|\,,
\end{multline}
where $z=\xi+i\eta$, $z'=\xi'+i\eta'$ and $\vartheta_{1}$ is the Jacobian theta function 
and reflects the double periodicity of the torus \cite{Theta}. A pedagogical derivation 
of the Green function $G_{0}(\bm{x},\bm{y})$ is reported in Appendix \ref{app:1}. 
Substituting Eq. \eqref{eq:periodic_green_function} in Eq. \eqref{eq:gamma} with 
$p_{2}=2\pi$ and
\[
p_{1} = 2\int_{0}^{\pi}\frac{d\psi}{r+\cos\psi} = \kappa\pi\,,
\]
we obtain:
\begin{equation}\label{eq:gamma_total}
\Gamma(\bm{x})=\frac{\pi}{3}\sum_{k=1}^{N}q_{k}\Gamma_{d}(\bm{x},\bm{x}_{k})-\Gamma_{s}(\bm{x})\,,
\end{equation}
where $\Gamma_{s}(\bm{x})$ represents the stress field due to the Gaussian curvature 
of the torus and is given by:
\begin{equation}\label{eq:gamma_screening}
\frac{\Gamma_{s}(\bm{x})}{Y}
= \log\left[\frac{r+\sqrt{r^{2}-1}}{2(r+\cos\psi)}\right]+\frac{r-\sqrt{r^{2}-1}}{r}\,.
\end{equation}
The function $\Gamma_{d}(\bm{x},\bm{x}_{k})$ is the stress field at the point 
$\bm{x}$ arising from the elastic distortion due to a defect at $\bm{x}_{k}$ and 
is given by
\begin{gather}
\frac{\Gamma_{d}(\bm{x},\bm{x}_{k})}{Y}
= \frac{\kappa}{16\pi^{2}}\left(\psi_{k}-\frac{2}{\kappa}\,\xi_{k}\right)^{2}-\frac{1}{4\pi^{2}\kappa}(\phi-\phi_{k})^{2}\notag\\[5pt]
+ \frac{1}{4\pi^{2}r}\log(r+\cos\psi_{k})-\frac{\kappa}{4\pi^{2}}\Real\{\Li(\alpha e^{i\psi_{k}})\}\notag\\[7pt]
+ \frac{1}{2\pi}\log\left|\vartheta_{1}\left(\frac{z-z_{k}}{\kappa}\bigg|\frac{2i}{\kappa}\right)\right|\,,
\label{eq:gamma_defects}
\end{gather}
where $\Li$ is the usual Eulerian dilogarithm and
\begin{equation}
\alpha = \sqrt{r^{2}-1}-r\,.
\end{equation}
A derivation of the functions $\Gamma_{s}(\bm{x})$ and $\Gamma_{d}(\bm{x},\bm{x}_{k})$ 
is given in Appendix \ref{app:2}. Integrating the function $\Gamma(\bm{x})$ on the 
torus gives the elastic energy of an arbitrary collection of disclinations. A 
detailed analysis of the crystalline structure arising from Eqs. \eqref{eq:free_energy}, 
\eqref{eq:gamma_total}, \eqref{eq:gamma_screening} and \eqref{eq:gamma_defects} is 
carried out in Sec. \ref{sec:3}. 

\section{\label{sec:3}Geometry of Toroidal Deltahedra}

Before analyzing the defect distribution arising from the elastic energy 
of Eq.~\eqref{eq:free_energy}, together with Eq.~\eqref{eq:gamma_total}, it
is necessary to understand the geometry of triangulated tori. Reconciling 
the predictions of a continuum elastic theory with the intrinsically 
discrete nature of crystallography requires an understanding of the 
possible lattices that can be embedded on the torus and the associated
defects. The problem of classifying the possible triangulations of the 
$2-$torus has received considerable attention from mathematicians, 
physicists and chemists over the past twenty years. Lavrenchenko 
\cite{Lavrenchenko:1990} proved in 1984 that all the triangulations of the
torus can be generated from 21 irreducible triangulations by certain 
sequences of operations called vertex splitting~\footnote{Analogously 
it can be proved that the number of irreducible triangulations 
is one for the sphere and two for the projective plane. The extraordinary 
larger value obtained for the torus should be indicative of the high 
structural complexity of a crystalline torus.}. After the discovery 
of carbon nanotubes in 1991 and the subsequent theoretical construction (later 
followed by the experimental observation) of graphitic tori, many possible
tessellations of the circular torus have been proposed by the community 
\cite{Dunlap:1992,ItohEtAl:1993,Kirby:1994,LaszloEtAl:2001,DiudeaEtAl:2001}.
In this section we review the construction of a defect-free triangulated 
torus and we show how the most symmetric defective triangulations can be 
generally grouped into two fundamental classes corresponding to symmetry 
groups $D_{nh}$ and $D_{nd}$ respectively. 

\begin{figure}
\includegraphics[width=.6\columnwidth]{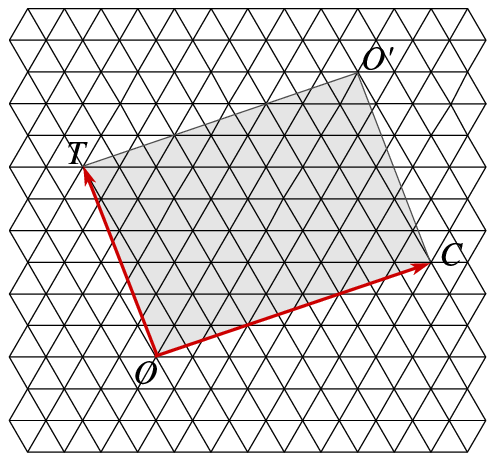}\\[7pt] 
\includegraphics[width=.6\columnwidth]{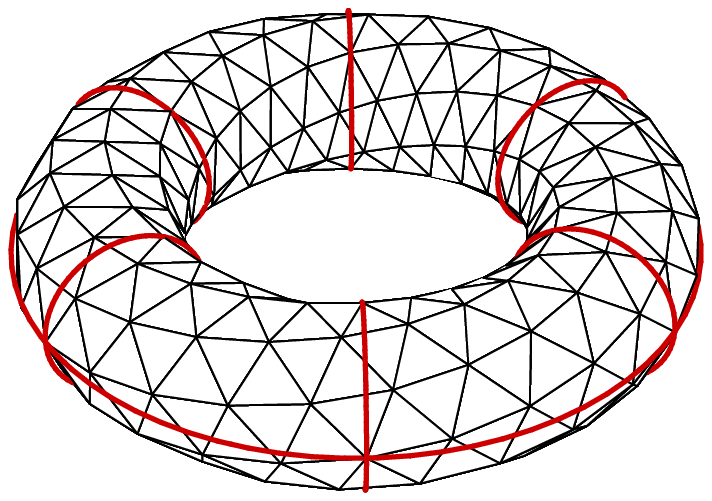}
\caption{\label{fig:triangulation}(Color online) Construction of a defect-free triangulation of 
the torus. On top planar map of the triangulated torus corresponding to the 
$(n,m,l)$ configuration $(6,3,1)$. On the bottom $(6,3,6)$ chiral torus. The edges 
of each one of the six tubular segments has been highlighted in red.}
\end{figure}

For the sake of consistency with the existing literature we adopt here the
language developed to describe the structure of carbon nanotubes. The 
structure of a triangulated cylinder can be specified by a pair of triangular
lattice vectors $\bm{c}$ and $\bm{t}$, called the \emph{chiral} and 
\emph{translation} vector respectively, which together define how the 
planar lattice is rolled up. In the canonical basis 
$\bm{a_{1}}=(1,0)$ and $\bm{a}_{2}=(\frac{1}{2},\frac{\sqrt{3}}{2})$, the 
vector $\bm{c}$ has the form:
\begin{equation}\label{eq:chiral_vector}
\bm{c} = n\bm{a}_{1}+m\bm{a}_{2}\qquad n,\,m\in\mathbb{Z}\,.
\end{equation}
The translation vector $\bm{t}$, on the other hand, can be expressed as 
an integer multiple 
\begin{equation}\label{eq:translation_vector}
\bm{t}=l\bm{e}_{t}\qquad l\in\mathbb{Z}
\end{equation}
of the shortest lattice vector $\bm{e}_{t}$ perpendicular to $\bm{c}$. The 
vector $\bm{e}_{t}$ is readily found to be of the form:
\[
\bm{e}_{t} = \frac{(n+2m)\bm{a}_{1}-(2n+m)\bm{a}_{2}}{(n+2m:2n+m)}\,,
\]
where $(a:b)$ denotes the greatest common divisor of $a$ and $b$ and 
enforces the minimal length. The three-dimensional structure of the torus 
is obtained by connecting the edge $\overline{OT}$ of the rectangle in Fig. 
\ref{fig:triangulation} (top) to $\overline{O'C}$ and $\overline{OC}$ to 
$\overline{O'T}$. The edge $\overline{OT}$ is then mapped to the external 
equator of the torus while the edge $\overline{OC}$ to the $\phi=0$ meridian. 
The resultant toroidal lattice has characteristic chirality related to the 
initial choice of the vector $\bm{c}$. In the nanotubes literature 
\emph{armchair} referes to the lattice obtained by choosing $n=m$, 
\emph{zigzag} to that obtained for $m=0$ and \emph{chiral} to all other 
lattices. An example of a $(n,m,l)$ chiral torus is shown in Fig.~\ref{fig:triangulation} 
(bottom) for the case $n=6$, $m=3$ and $l=6$. The chirality is extremely important
in graphitic carbon nanotube or nanotori, where it determines whether
the electronic behavior of the system is metallic or semiconducting.

By Euler's theorem one can prove that the number of triangular faces $F$ 
and the number of vertices $V$ of a triangular toroidal lattice is given by:
\[
V = \tfrac{1}{2}F\,.
\]
Denoting $A_{R}$ the area of the rectangle with edges $\bm{c}$ and $\bm{t}$ 
and $A_{T}$ the area of a fundamental equilateral triangle, the number of
vertices of a defect-free toroidal triangulation is then:
\begin{equation}\label{eq:vertices}
V 
= \frac{A_{R}}{2A_{T}}
= \frac{2l\,(n^{2}+nm+m^{2})}{(n+2m:2n+m)}\,.
\end{equation}
The planar construction reviewed above allows only lattices with an even 
number of vertices. Defect-free toroidal deltahedra with an odd number of 
vertices are also possible and their construction is generally achieved by 
assembling congruent octahedral building blocks. An example of this scheme 
will be briefly discussed in Sec.~\ref{sec:5} for the case $V=87$ and $r=6$.
We refer the reader to Ref. \onlinecite{Webber:1997} for an comprehensive review
of the topic.

\begin{figure}
\centering
\includegraphics[width=.9\columnwidth]{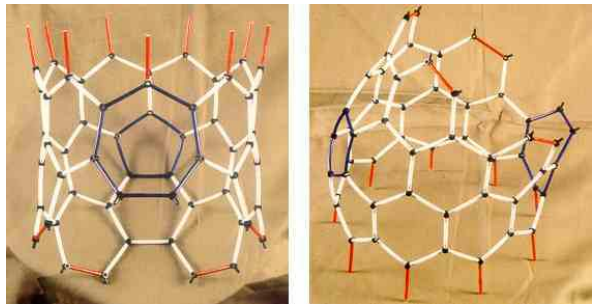}
\caption{\label{fig:knees}(Color online) Dunlap knees obtained by joining two straight tubular
segments with $(n,0)$ and $(m,m)$ chirality. [Courtesy of A. A. Lucas and 
A. Fonseca, Facult\'es Universitaries Notre-Dame de la Paix, Namur, Belgium].}
\end{figure}

\begin{figure}
\centering
\includegraphics[width=.9\columnwidth]{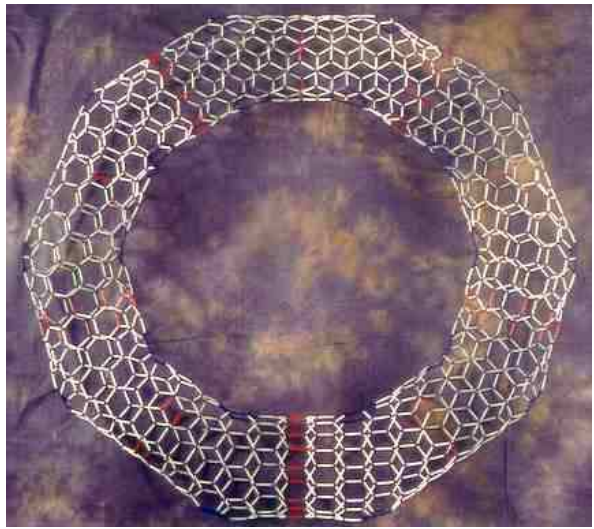}
\caption{\label{fig:lucas-torus}(Color online) Five-fold polygonal torus obtained by joining
$(5,5)$ and $(9,0)$ tubular segments via ten pairs of $5-7$ rings. This structure
was originally proposed by the authors of Ref.~\onlinecite{LambinEtAl:1995} as 
a possible low-strain configuration for carbon nanotori. [Courtesy of A. A. Lucas 
and A. Fonseca, Facult\'es Universitaries Notre-Dame de la Paix, Namur, Belgium].}
\end{figure}

The embedding of an equal number of pentagonal and heptagonal disclinations 
in the hexagonal network was first proposed by Dunlap in 1992 as a possible 
way to incorporate positive and negative Gaussian curvature into the cylindrical 
geometry of carbon tubules \cite{Dunlap:1992}. According to the Dunlap 
construction the necessary curvature is incorporated by the insertion of 
``knees'' (straight cylindrical sections of the same diameter joined with a
kink) in correspondence with each pentagon-heptagon pair arising from the 
junction of tubular segments of different chirality (see Fig.~\ref{fig:knees}).   
\begin{figure}[b]
\centering
\includegraphics[width=0.7\columnwidth]{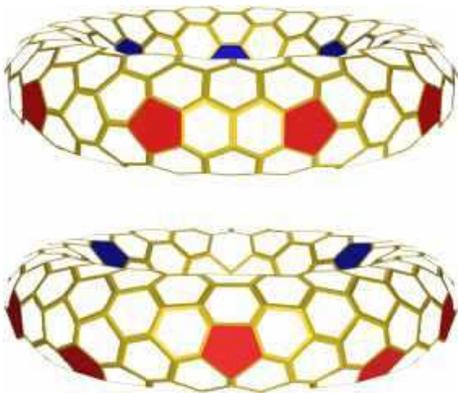}
\caption{\label{fig:prism}(Color online) Voronoi lattices of a TP$n$ prismatic (top) and 
TA$n$ antiprismatic (bottom) toroids with $R_{1}=1$ and $R_{2}=0.3$.}
\end{figure}
In particular, a junction between a $(n,0)$ and a $(m,m)$ tube can be obtained  
by placing a $7-$fold disclination along the internal equator of the torus and 
a $5-$fold disclination along the external equator. Since the radii of the two 
segments of a junction are different by construction, the values of $n$ and $m$ 
are commonly chosen to minimize the ratio $|\bm{c}_{(n,0)}|/|\bm{c}_{(m,m)}|=n/\sqrt{3}m$.
By repeating the $5-7$ construction periodically it is possible to construct an 
infinite number of toroidal lattices with an even number of disclinations pairs 
and dihedral symmetry group $D_{nh}$ (where $2n$ is the total number of $5-7$ pairs,
Fig~\ref{fig:lucas-torus}). The structure of the lattice is described by the 
alternation of two motifs with crystalline axes mutually rotated by $30^{\circ}$ 
as a consequence of the connecting disclination. One of the fundamental aspects 
of Dunlap's construction is that all the disclinations are aligned along the two 
equators of the torus where the like-sign Gaussian curvature is maximal. As we will 
see below, this feature makes these arrangements optimal in releasing the elastic 
stress due to curvature.

Another class of crystalline tori with dihedral antiprismatic symmetry $D_{nd}$ was 
initially proposed by Itoh \emph{et al} \cite{ItohEtAl:1993} shortly after Dunlap. 
Aimed at reproducing a structure similar to the C$_{60}$ fullerene, Itoh's 
original construction implied ten disclination pairs and the point group $D_{5d}$. 
In contrast to Dunlap tori, disclinations are never aligned  along the equators in antiprismatic tori,  
instead being staggered at some angular distance $\delta\psi$ from the 
equatorial plane. Hereafter we will use the symbol TA$n$ to refer to toroidal deltahedra
with $2n$ disclination pairs and $D_{nd}$ symmetry group.

\begin{figure}[t]
\centering
\includegraphics[width=.9\columnwidth]{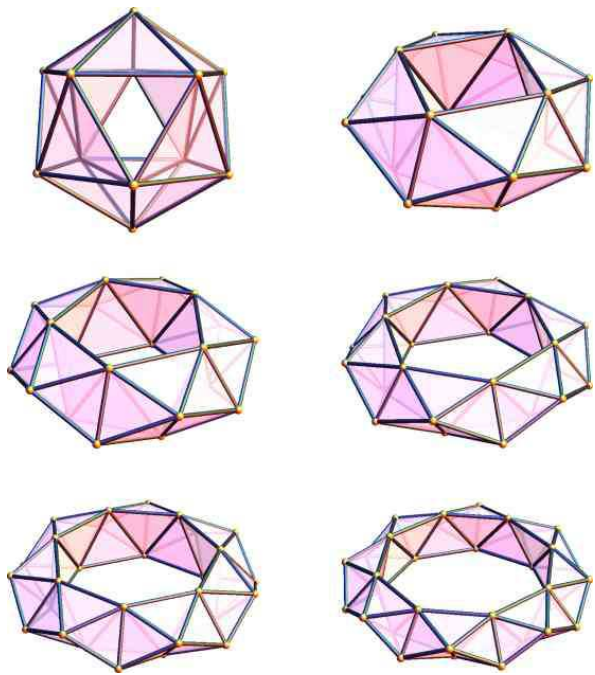}
\caption{\label{fig:toroidal-antiprisms}(Color online) First six toroidal antiprisms obtained
by repeating the unit cell of Fig.~\ref{fig:graph1}. The first polyhedron on 
the left is the ``drilled icosahedron''.}
\end{figure}

\begin{figure}[h]
\centering
\includegraphics[scale=0.4]{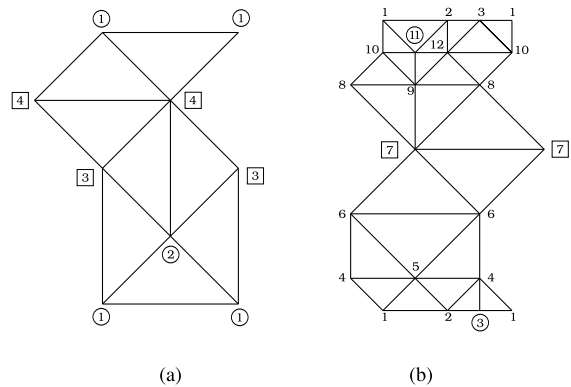}
\caption{\label{fig:graph1}(a) Unit cell for toroidal antiprisms. $5-$fold vertices
are circled and $7-$fold vertices are boxed. (b) Unit cell of a $D_{nd}$ torus in the 
Berger-Avron construction. The graph consists of four generation of tiles and the 
internal equator of the torus is mapped into the horizontal line passing to the 
mid-point between the 6th and the 7th vertex.}
\end{figure}

A systematic construction of defected triangulations of the torus can be 
achieved in the context of planar graphs \cite{Lavrenchenko:1990,BergerAvron:1995}.
A topological embedding of a graph in a two-dimensional manifold corresponds to a 
triangulation of the manifold if each region of the graph is bounded by exactly 
three vertices and three edges, and any two regions have either one common vertex 
or one common edge or no common elements of the graph. The simplest example of 
toroidal polyhedra with $D_{nd}$ symmetry group, featuring only $5-$fold and 
$7-$fold vertices, can be constructed by repeating $n$ times the unit cell of
Fig.~\ref{fig:graph1}a. These \emph{toroidal antiprisms}~\footnote{Although we 
presume this class of toroidal polyhedra is not discussed here for the first 
time, we couldn't find any previous reference in the literature.} have 
$V=4n$ vertices and can be obtained equivalently from the edge skeleton of a 
$n-$fold antiprism by attaching at each of the base edges a pentagonal pyramid 
and by closing the upper part of the polyhedron with $n$ additional triangles. 
By counting the faces one finds $F=5n+2n+n=8n$ from which $V=4n$. The simplest 
polyhedron of this family has $V=12$ and $D_{3d}$ symmetry group (see top left of Fig.
\ref{fig:toroidal-antiprisms}) and corresponds to the ``drilled icosahedron''
obtained by removing two parallel faces of an icosahedron and connecting the
corresponding edges with the six lateral faces of an antiprism with triangular
base (i.e. a prolate octahedron). Starting from this family of toroidal antiprisms
a number of associated triangulations having the same defect structure can be 
obtained by geometrical transformations such as the Goldberg inclusion \cite{Goldberg:1937,CasparKlug:1962,
VirusMacromolecules}. Such transformations, popularized by Caspar and Klug for the 
construction of the icosadeltahedral structure of spherical viruses \cite{CasparKlug:1962}, 
consist in partitioning each triangular face of the original graph into smaller 
triangular faces in such a way that old vertices preserve their valence and new 
vertices have valence six. The partition is obtained by specifying two integer numbers 
$(L,M)$ which define how the original vertices of each triangle are connected by the 
new edges so that the total number of vertices is increased by a factor $T = L^{2}+LM+M^{2}$. 

A general classification scheme for $D_{nd}$ symmetric tori was provided by Berger 
and Avron \cite{BergerAvron:1995} in 1995. Their scheme is based on the construction 
of unit graphs comprising triangular tiles of different \emph{generations}. In each 
generation, tiles are scaled in length by a factor $1/\sqrt{2}$ with respect to the 
previous generation. This rescaling approximates the non-uniformity of 
the metric of a circular torus.

\begin{figure}[b]
\centering
\includegraphics[scale=0.4]{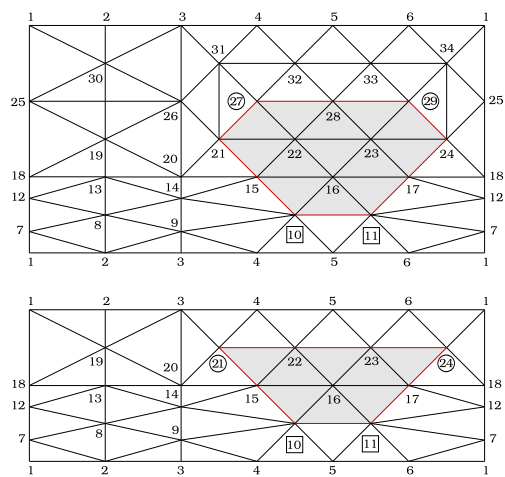}
\caption{\label{fig:graph2}(Color online) Unit cells for Dunlap toroids of type $(2,1,3,1)$ and
$(3,1,3,1)$ according to the classification scheme given here. Highlighted 
regions correspond to the central polygon.}
\end{figure}

\begin{figure}[t]
\centering
\includegraphics[width=0.9\columnwidth]{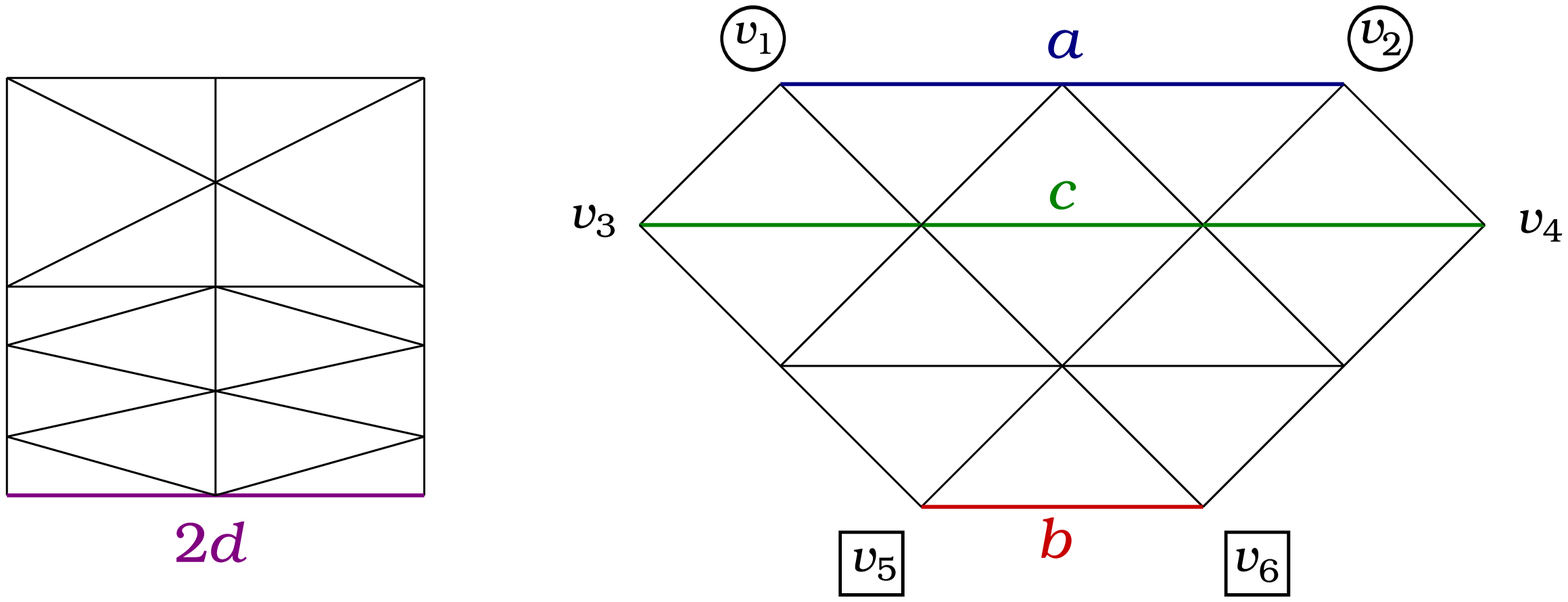}
\caption{\label{fig:graph3}(Color online) Central polygon (right) and rectangular (zig-zag)
region in our construction scheme of Dunlap's toroids. In this example $(a,b,c,d)=(2,1,3,1)$.}
\end{figure}

Dunlap toroids can be obtained from unit cells such as those shown in 
Fig.~\ref{fig:graph2}. The geometrical properties of these graphs can
be described in different ways. A particularly intuitive way, in the 
spirit of this paper, consists in specifying the distances between 
$5-$ and $7-$fold pairs. One starts by drawing the smallest convex
loop passing through defective sites. This identifies a central 
polygon whose upper vertices ($v_{1}$ and $v_{2}$ in Fig.\ref{fig:graph3}) 
have degree five and lower vertices ($v_{5}$ and $v_{6}$ in Fig.~\ref{fig:graph3}) 
have degree seven. Then calling $a$ the distance between $5-$fold 
vertices $v_{1}$ and $v_{2}$, $b$ that between $7-$fold vertices 
$v_{5}$ and $v_{6}$ and $c$ the length of the segment $\overline{v_{3}v_{4}}$
($d=a$ for a trapezoid), we can express the total number of triangles 
enclosed by the central polygon as:
\[
f=2c^{2}-a^{2}-b^{2}\,.
\]
Each $7-$fold vertex sits at the apex of a diamond-shaped complex of 
$f'=7$ triangles. Each $5-$fold vertex, on the other hand, is at the 
apex of a triangular region of $f''=(c-a+1)^{2}$ triangles. The graph 
is completed by a rectangle of height $c-a+4$ and base of arbitrary
length $2d$ containing:
\[
f'''= 4d(c-a+4)
\]
triangles. The total number of vertices is
\begin{align}\label{eq:dunlap-graphs}
V_{g}
&=f+f'+f''+f'''/2\notag\\
&=c^{2}-b^{2}+2(c-a)(c+d+1)+8(d+1)\,.
\end{align}
The final triangulation of the torus is obtained by repeating the prismatic 
unit cell $l$ times and therefore has $V=lV_{g}$ vertices. This scheme 
provides direct information on the arrangement of defective sites. 
Thus for instance an $(a,b,c,d)=(2,1,3,1)$ unit 
cell (see top of Fig.~\ref{fig:graph2}) has $5-$fold vertices separated by two lattice spacings   
and $7-$fold vertices by one lattice spacing. On the other hand the integers $n$ and $m$ giving the chirality
of the two segments of the junction $(n,0)/(m,m)$ are given directly by
as:
\begin{align*}
n &= c-a+4 \\
m &= 2c-a-b\, .
\end{align*}
Thus the $(2,1,3,1)$ cell of Fig.~\ref{fig:graph2} is obtained from the 
junction between a $(5,0)$ and a $(3,3)$ tubular segment.

\begin{figure}[h]
\centering
\includegraphics[scale=0.4]{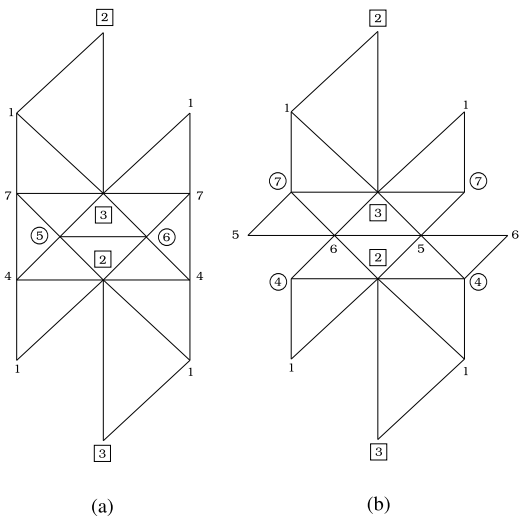}
\caption{\label{fig:graph4}Unit cells for TP(2)$n$ and TP(3)$n$ 
toroids.}
\end{figure}

\begin{figure}
\centering
\includegraphics[width=.6\columnwidth]{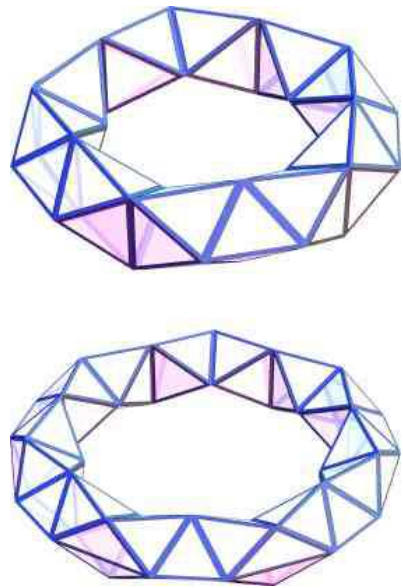}
\caption{\label{fig:prismatic-deltahedra}(Color online) (a) TP5a and (b) TP7b 
toroids with $V=35$ and 49 obtained by repeating the unit cells of
Fig.~\ref{fig:graph4}. $7-$fold vertices form dimers normal to 
the equatorial plane while $5-$fold vertices are (a) distributed
along the external equator or (b) form a double ring above and
below the equatorial plane.}
\end{figure}

Dunlap's toroids are not the only examples of defective triangulations
of the torus with dihedral prismatic symmetry group $D_{nh}$. With the 
help of numerical simulations (see Sec.~\ref{sec:5}) we found two other
classes whose unit cell is shown in Fig.~\ref{fig:graph4}. Unlike Dunlap's 
toroids, the $7-$fold vertices in these prismatic triangulations are not aligned 
along the internal equator of the torus, but rather grouped in dimers 
normal to the equatorial plane. $5-$fold vertices are distributed
along the external equator in the graph of Fig.~\ref{fig:graph4}a 
or form a double ring above and below it in the case of the graph
Fig.~\ref{fig:graph4}b. Toroidal deltahedra obtained by embedding the
prismatic graphs of Fig.~\ref{fig:graph4} on a circular torus are 
shown in Fig.~\ref{fig:prismatic-deltahedra} for the case of a 
$5-$fold symmetric toroid with $V=35$ and a $7-$fold symmetric 
toroid with $V=49$. In the rest of the paper we will reserve the
symbol TP$n$ for Dunalp's toroids and refer with TP$n$a and 
TP$n$b to the other two classes of toroids with symmetry group
$D_{nh}$ and unit cell of as shown in Fig.~\ref{fig:graph4}a and 
Fig.~\ref{fig:graph4}b respectively. 

\begin{figure}
\centering
\includegraphics[width=0.7\columnwidth, clip]{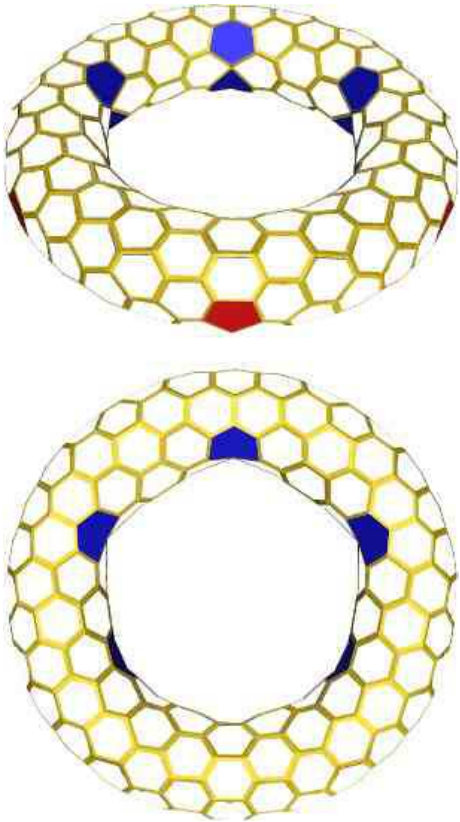}
\caption{\label{fig:r_3.3_200}(Color online) Voronoi diagram of a toroidal lattice with $r=10/3$ 
and $V=200$ vertices. The lattice exhibits 11 disclination pairs and has $C_{s}$ 
symmetry group.}
\end{figure}

All defective triangulations presented so far are characterized by an even number
of disclination pairs. Regular tessellations of the torus comprising an odd number 
of defects pairs are also possible. Such tessellations are obtained by combining 
segments of prismatic and antiprismatic lattices with a consequent loss of dihedral 
symmetry. Fig. \ref{fig:r_3.3_200} shows the Voronoi diagram of a toroidal lattice, 
with $r=10/3$ and $V=200$ vertices, containing 11 disclination pairs. For an angular 
length of approximately $\Delta\phi=7/5\,\pi$ the lattice is a prismatic $D_{5d}$ 
toroid while in the remaining $3/5\,\pi$ the local structure is that of an 
antiprismatic toroid. The global structure has only bilateral symmetry about a 
sagittal plane dividing the lattice in two mirror halves and thus point group 
$C_{s}$.

In the past few years, alternative constructions of triangulated tori have been 
proposed as well as novel geometrical and graph-theoretical methods to express 
the coordinates of their three-dimensional structures (see for example Kirby 
\cite{Kirby:1994}, L\'{a}szl\'{o} \emph{at al} \cite{LaszloEtAl:2001}, Diudea 
\emph{et al} \cite{DiudeaEtAl:2001}). Here we choose to focus on the defect 
structure associated with the two most important class TP$n$ and TA$n$ with
groups $D_{nh}$ and $D_{nd}$.

\section{\label{sec:4}Disclinations and Scars}

\subsection{\label{sec:isolated}Isolated Defects Regime}

In this section we analyze the crystalline structures arising from the solution 
of the elastic problem and we show how the interplay between the geometry of the 
embedding torus, the topology of the lattice, and the mechanical properties of 
the microscopic units, here encoded in the Young modulus $Y$ and the core energy, 
lead to a rich variety of structures whose phase-diagram is presented at the 
end of the section. Since the free energy of Eq.~\eqref{eq:free_energy} is 
minimized when disclinations best approximate the continuum Gaussian curvature
of the torus, it is clear that disclinations are most likely to be found in 
regions of like-sign Gaussian curvature. Maximal curvature occurs along the 
external (positive) and internal (negative) equators, which thus constitute 
preferred regions for the appearance of disclinations in the ground state.

\begin{figure}[b]
\centering
\includegraphics[width=1.\columnwidth]{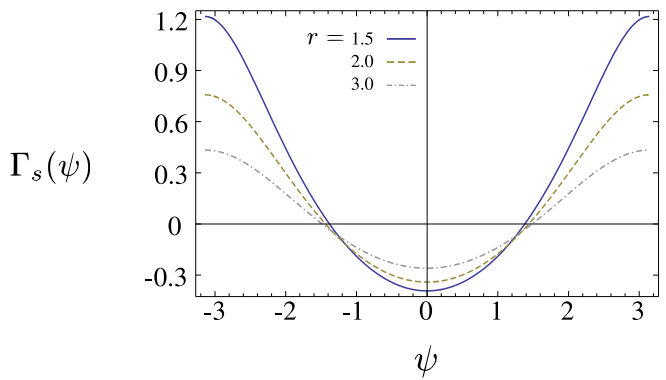}
\caption{\label{fig:gamma_s}(Color online) The screening function $\Gamma_{s}(\psi)$ for 
different values of the aspect ratio $r$.}
\end{figure}     

\begin{figure}[t]
\centering
\includegraphics[width=1\columnwidth]{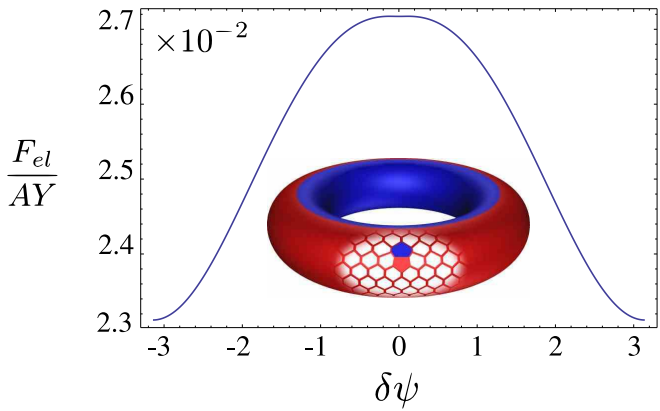}
\caption{\label{fig:redbluetorus}(Color online) Elastic energy of a $5-7$
disclination dipole constrained to lie on the same meridian, as a
function of the angular separation. In the inset, illustration of a
circular torus of radii $R_{1}>R_{2}$. Regions of positive and
negative Gaussian curvature have been shaded in red and blue
respectively.}
\end{figure}

\begin{figure}[b]
\centering
\includegraphics[width=1\columnwidth]{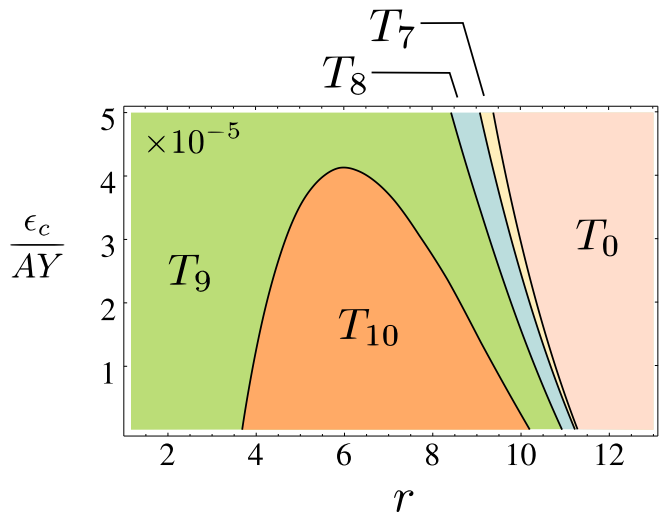}
\caption{\label{fig:phase_diagram1}(Color online) Phase diagram for $T_{p}$ configurations 
in the plane $(r,\epsilon_{c}/AY)$. For $r\in[3.68,\,10.12]$ and 
$\epsilon_{c}\sim 0$ the structure is given by a $T_{10}$ configuration with 
symmetry group $D_{5h}$.}
\end{figure}

To analyze the elastic free energy \eqref{eq:free_energy} we start by considering 
the energies of two opposite sign disclinations constrained to lie on the same 
meridian. The elastic free energy of this system is shown in  
Fig.~\ref{fig:redbluetorus} as a function of the angular separation between the 
two disclinations. The energy is minimized for the positive ($5-$fold) disclination
on the external equator (maximally positive Gaussian curvature) and the negative 
($7-$fold) disclination on the internal equator (maximally negative Gaussian 
curvature). The picture emerging from this simple test case suggests that a
good \emph{ansatz} for an optimal defect pattern is a certain number $p$ of
equally spaced $+1$ disclinations on the external equator matched by the same
number of equally spaced $-1$ disclinations on the internal equator.
We name this configuration with the symbol $T_{p}$, where $p$ stands for the 
total number of disclination pairs.
\begin{equation}\label{eq:s_n}
T_{p}:\quad
\left\{
\left( 0, \frac{2\pi k}{p}\right)_{1 \le k \le p};\,
\left(\pi,\frac{2\pi k}{p}\right)_{1 \le k \le p}
\right\}\,,
\end{equation}
where the two pairs of numbers specify the $(\psi,\phi)$ coordinates of the 
positive and negative disclinations respectively. A comparison of the energy 
of different $T_{p}$ configurations, as a function of aspect ratio and disclination
core energy, is summarized in the phase diagram of Fig. \ref{fig:phase_diagram1}. 
We stress here that only $T_{p}$ configurations with $p$ even have an embedding on 
the torus corresponding to lattices of the TP$\frac{p}{2}$ class. Nevertheless a 
comparison with $p-$odd configurations can provide additional information on the 
stability of $p-$even lattices. For small core energies, moreover, thermally 
excited configurations with a large number of defects and similar $p-$polar 
distributions of topological charge are expected to exhibit an elastic energy 
comparable in magnitude with that of these minimal constructions.      
The defect core energy has been expressed here in the form:
\begin{equation}\label{eq:core_energy}
F_{c} 
= \epsilon_{c}\sum_{i=1}^{2p} q_{i}^{2} 
= 2p\epsilon_{c} \ .
\end{equation}
The core energy $\epsilon_{c}$ of a single disclination depends on the details of 
the crystal-forming material the corresponding microscopic interactions. A simple 
phenomenological argument (see for example Ref.~\onlinecite{KlemanLavrentovich}) 
gives
\[
\frac{\epsilon_{c}}{Y} \sim \frac{a^{2}}{32\pi} \ ,
\]
where $a$ the lattice spacing. Taking $a^{2}=A/\frac{\sqrt{3}}{2}V$, with $A$ the 
area of the torus, yields:
\begin{equation}\label{eq:core_magnitude}
\frac{\epsilon_{c}}{AY} \sim \frac{1}{16\sqrt{3}\,\pi V} \sim \frac{10^{-2}}{V}\,.
\end{equation}
For a system of order $V=10^{3}$ subunits, then, the dimensionless core energy on
the left hand side of Eq.~\eqref{eq:core_magnitude} is of order $10^{-5}$. This
estimate motivates our choice of the scale for $\epsilon_{c}/(AY)$ in Fig. 
\ref{fig:phase_diagram1}. 

For dimensionless core energies below $4\cdot 10^{-5}$ and aspect ratios $r$
between 3.68 and 10.12 the ground state structure is the TP5 lattice 
corresponding to a double ring of $+1$ and $-1$ disclinations distributed on
the external and internal equators of the torus as the vertices of a regular 
decagon (the $T_{10}$ configuration). The TP5 lattice has dihedral symmetry group
$D_{5h}$. That this structure might represent a stable configuration for polygonal 
carbon toroids has been conjectured by the authors of Ref. 
\onlinecite{LambinEtAl:1995}, based on the argument that the 36$^{\circ}$ angle 
arising from the insertion of ten pentagonal-heptagonal pairs into the lattice 
would optimize the geometry of a nanotorus consistently with the structure of the 
$sp^{2}$ bonds of the carbon network (unlike the 30$^{\circ}$ angle of the $6-$fold 
symmetric configuration originally proposed by Dunlap). In later molecular
dynamics simulations, Han \cite{Han:1997} found that a $5-$fold symmetric 
lattice, such as the one obtained from a (9,0)/(5,5) junction (see Fig.~\ref{fig:lucas-torus}), 
is in fact stable for toroids with aspect ratio less then $r\sim 10$. The stability, 
in this case, results from the strain energy per atom being smaller than 
the binding energy of carbon atoms. Irrespective of the direct experimental 
observation of such disclinated toroidal crystals, which is still open, we have 
shown here, from continuum elasticity, that a  $5-$fold symmetric lattice indeed 
constitutes a minimum of the elastic energy for a broad range of aspect ratios 
and defect core energies.

\begin{figure}[t]
\centering
\includegraphics[width=0.55\columnwidth]{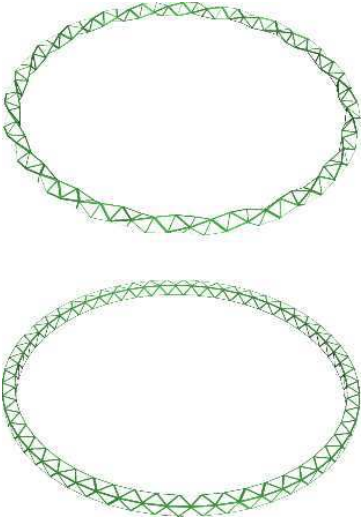}
\caption{\label{fig:cropcircle}(Color online) Two examples of defect free ``crop circle'' 
toroids with $r=20$ and $V=180$ (left) and $220$ (right).}
\end{figure}

\begin{figure}[b]
\centering
\includegraphics[width=1\columnwidth]{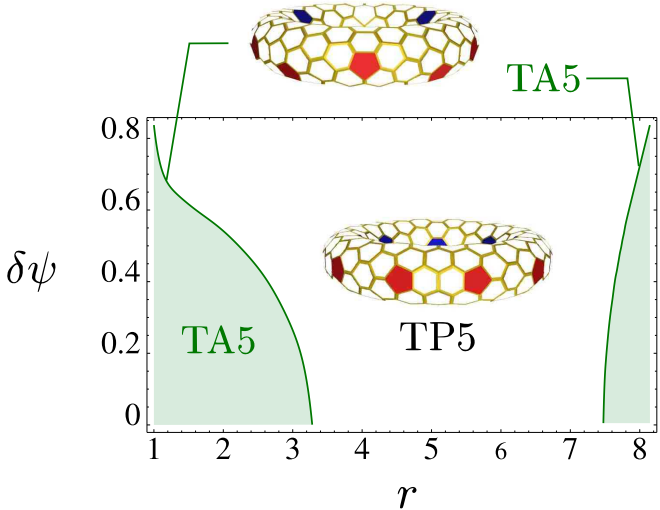}
\caption{\label{fig:phase_diagram2}(Color online) Phase diagram of a $5-$fold
symmetric lattice in the plane $(r,\delta\psi)$. For small
$\delta\psi$ and $r$ in the range $[3.3,\,7.5]$ the prismatic TP5
configuration is energetically favored. For $r<3.3$ the system
undergoes a structural transition to the antiprismatic phase TA5.}
\end{figure}

For small aspect ratios the $5-$fold symmetric configuration becomes 
unstable and is replaced by the $9-$fold symmetric phase $T_{9}$. As we 
mentioned, however, this configuration doesn't correspond to a possible 
triangulation of the torus. It is likely that the ground sate in this regime
consists of ten skew disclination pairs as in the antiprismatic TA$n$ 
lattice. The latter can be described by introducing a further degree of 
freedom $\delta\psi$ representing the angular displacement of defects from 
the equatorial plane:
\begin{multline}
\mathrm{TA}n:\quad
\left\{
\left( (-1)^{2k}\delta\psi, \frac{2\pi k}{n}\right)_{1 \le k \le n};\right.\\[7pt]
\left.\left( (-1)^{2k}(\pi-\delta\psi),\frac{2\pi k}{n}\right)_{1 \le k \le n}
\right\}
\end{multline}
A comparison of the TP5 configuration and the TA5 configuration is 
shown in Fig.~\ref{fig:phase_diagram2} for different values of 
$\delta\psi$. The intersection points of the boundary curves with the 
$\delta\psi-$axis has been calculated by extrapolating the $(r,\delta\psi)$ 
data points in the range $\delta\in[0.07,\,0.8]$ with 
$\Delta(\delta\psi) = 2.5\pi\cdot 10^{-3}$. For small $\delta\psi$ and 
$r\in[3.3,\,7.5]$ the prismatic TP5 configuration is energetically favored. 
For $r<3.3$, however, the lattice undergoes a structural transition to the TA5 
phase. For $r>7.5$ the prismatic symmetry of the TP5 configuration breaks
down again. In this regime, however, the elastic energy of both configurations 
rapidly rises because of the lower curvature and defects disappear.

\subsection{\label{sec:defect-free}Defect-free Tori}

For aspect ratio $r\gtrsim 11$ the TP5 phase is replaced by a defect free 
configuration ($T_{0}$ in Fig.~\ref{fig:phase_diagram1}) so that
configurations with defects are no longer energy minima. Any toroidal crystal 
with aspect ratio larger than $\sim 11$ is than energetically favored to be 
defect-free. In the thin torus limit the ground state structure is directly 
related to the simple problem of finding the most efficient 
packing of congruent equilateral triangles on the torus of a given aspect 
ratio. Given $V$ subunits (vertices) one seeks the densest packing of equilateral 
triangles of edge-length $a=(A/\frac{\sqrt{3}}{2}\,V)^{1/2}$ on the torus with 
aspect ratio $r$, such that each vertex has valence six. Using the planar 
construction described in Sec. \ref{sec:3}, the optimal choice of the indices 
$(n,m,l)$, can be translated into the minimization of the following quantity:
\begin{equation}\label{eq:chirality_condition}
\Delta^{n,m}(r,V) 
=n^{2}+nm+m^{2}-\tfrac{\sqrt{3}}{2}\,r^{-1}V\,,
\end{equation}
obtained by equating the magnitude of the chiral vector $\bm{c}$ with that of 
the sectional circumference of the embedding torus, under the constraints: 
\begin{equation}
\left\{
\begin{array}{l}
l = \frac{V}{2}\frac{(n+2m:2n+m)}{n^{2}+nm+m^{2}}\\[7pt]
n,\,m,\,l\in\mathbb{Z}
\end{array}
\right.\,.
\end{equation}
This construction successfully predicts the structure of the lattices of 
Fig.~\ref{fig:cropcircle}. 

\begin{figure}
\centering
\includegraphics[width=1.\columnwidth]{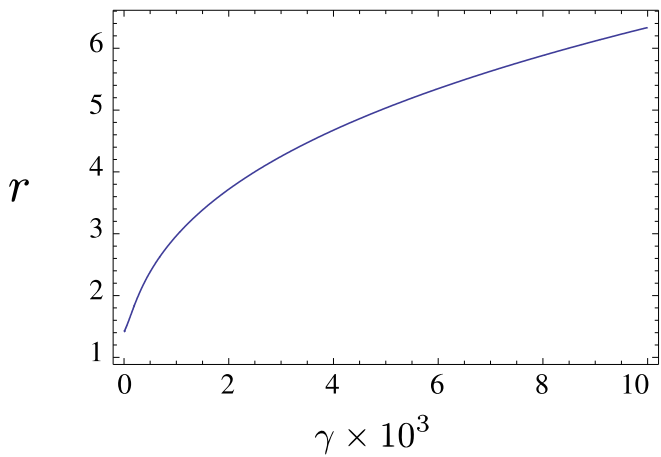}
\caption{\label{fig:optimalr}(Color online) Optimal value of the
aspect ratio $r$ as function of the F\"oppl-von K\'arm\'an number 
$\gamma=AY/\kappa_{b}$. For $\gamma\sim 0$ the Clifford torus 
with $r=\sqrt{2}$ is optimal. Larger values of $\gamma$ favour
instead a ``skinnier'' torus.}
\end{figure}

So far we have studied the elasticity of toroidal crystals exclusively in
terms of interacting topological defects on a rigid toroidal substrate. 
Thus the elastic strain due to defects and curvature takes the form of pure
stretching on the tangent plane of the torus and no out-of-plane deformation 
takes place. In a more realistic scenario,
a crystalline torus would undergo both in-plane stretching and out-of-plane
bending. The latter implies an energy cost:
\begin{equation}\label{eq:bending}
F_{b} 
= \frac{\kappa_{b}}{2} \int d^{2}x\, H^{2}(\bm{x})
= \kappa_{b}\frac{2\pi^{2}r^{2}}{\sqrt{r^{2}-1}} \ ,
\end{equation}
with $H$ the mean curvature (see Ref.~\onlinecite{BowickNelsonTravesset:2004}). 
The case of defect-free tori is simple enough to incorporate bending in the problem 
and see what the optimal aspect ratio of a defect-free torus would be as a
function of the F\"oppl-von K\'arm\'an number $\gamma=AY/\kappa_{b}$
representing the ratio of the stretching energy scale to the 
bending rigidity. In absence of defects the only source of stress
is given by the curvature. Thus
\begin{align}
F_{s} 
&= \frac{1}{2Y} \int d^{2}x\,\Gamma_{s}^{2}(\bm{x})\notag\\
&= AY\Big\{\frac{1+4r(r^{2}-1)^{\frac{1}{2}}[1-\log(2+2r\alpha)]}{2r^{2}}\notag\\
&\hspace{0.47\columnwidth}+\Li(\alpha^{2})-2\Big\}\,.\label{eq:stretching}
\end{align}
Summing Eq.~\eqref{eq:bending} and \eqref{eq:stretching} and taking
the derivative with respect to $r$ (assuming constant area), one obtains 
the following equation for the optimal value of
$r$:
\begin{multline}\label{eq:optimalr}
2\pi^{2}\frac{r(r^{2}-2)}{(r^{2}-1)^{\frac{3}{2}}}
-\frac{\gamma}{r^{3}}\Big[1+2r\alpha\\-2r(r^{2}-1)^{\frac{1}{2}}\log(2+2r\alpha)\Bigg]=0\,.
\end{multline}
The optimal aspect ratio $r$ as obtained form Eq.~\eqref{eq:optimalr}
is shown in Fig.~\ref{fig:optimalr} as a function of $\gamma$. For 
$\gamma\sim 0$, when the major contribution to the elastic energy 
is given by the bending, the optimal geometry is given by the 
Clifford torus ($r=\sqrt{2}$). If, on the other hand, the in-plane
stretching dominates, a ``skinny'' torus (large $r$) is energetically
favoured.

\subsection{\label{sec:scars}The Coexistence Regime}

In this section we investigate the possibility of a regime of coexistence 
between isolated disclinations and grain boundary ``scars''. The existence 
of scars, first predicted in the context of spherical crystallography 
\cite{BowickNelsonTravesset:2000} and later observed experimentally in 
spherical droplets coated with colloids \cite{BauschEtAl:2003,EinertEtAl:2005}, has 
become one of the fundamental signatures of dense geometrically frustrated 
systems.

In the regime of large particle numbers, the amount of curvature
required to screen the stress field of an isolated disclination in
units of lattice spacing becomes too large and disclinations are
unstable to grain boundary ``scars'' consisting of a linear array of
tightly bound $5-7$ pairs radiating from an unpaired disclination
\cite{BowickNelsonTravesset:2000,GiomiBowick:2007}. In a manifold
with variable Gaussian curvature this effects leads to a regime of 
coexistence of isolated disclinations (in regions of large curvature) 
and scars. In the case of the torus the Gaussian curvature inside 
($|\psi|>\pi/2$) is always larger in magnitude than that outside 
($|\psi|<\pi/2$) for any aspect ratio and so we may expect a regime 
in which the negative internal curvature is still large enough to 
support the existence of isolated $7-$fold disclinations, while on 
the exterior of the torus disclinations are delocalized in the form 
of positively charged grain boundary scars.

\begin{figure}[b]
\centering
\includegraphics[width=1\columnwidth]{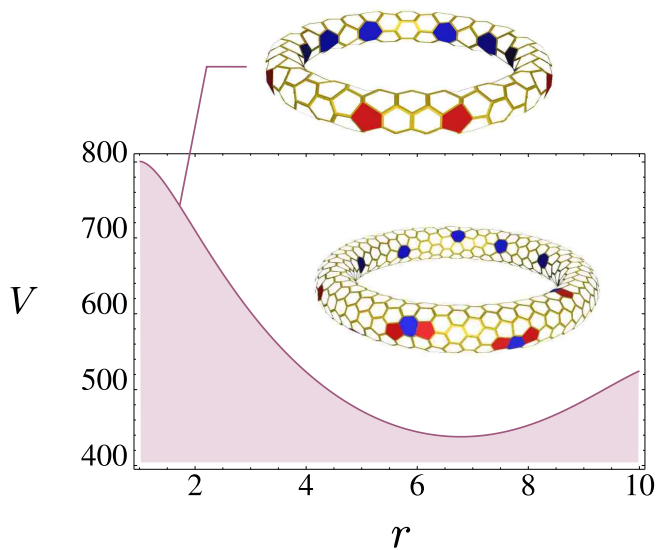}
\caption{\label{fig:phase_diagram3}(Color online) Isolated defects and 
scar phases in the $(r,V)$ plane. When the number of vertices $V$ 
increases the range of the screening curvature becomes smaller than 
one lattice spacing and disclinations appear delocalized in the form 
of a $5-7-5$ grain boundary mini-scar.}
\end{figure}

To check this hypothesis we compare the energy of the TP5 lattice
previously described with that of ``scarred'' configurations
obtained by decorating the original toroid in such a way that each
$+1$ disclination on the external equator is replaced by a
$5-7-5$ mini-scar. The result of this comparison is summarized in the
phase diagram of Fig. \ref{fig:phase_diagram3} in terms of $r$ and
the number of vertices of the triangular lattice $V$ (the
corresponding hexagonal lattice has twice the number of vertices,
i.e. $V_{hex}=2V$). $V$ can be derived from the angular separation
of neighboring disclinations in the same scar by approximating $V
\approx A/A_{V}$, with $A_{V}=\frac{\sqrt{3}}{2}a^{2}$ the area of a
hexagonal Voronoi cell and $a$ the lattice spacing. When the aspect
ratio is increased from 1 to 6.8 the range of the curvature
screening becomes shorter and the number of subunits required to
destroy the stability of the TP5 lattice decreases. For $r>6.8$,
however, the geodesic distance between the two equators of the torus
becomes too small and the repulsion between like-sign defects takes
over. Thus the trend is inverted.

\section{\label{sec:5}Numerical Experiments}

In this section we report the result of a numerical minimization of
a system of $V$ point-like particles constrained to lie on the surface
of a torus and interacting via a pair potential of the form 
$U_{ij}=1/|\bm{x}_{i}-\bm{x}_{j}|^{3}$ where $|\cdot|$ denotes the
Euclidean distance in $\mathbb{R}^{3}$. The problem of finding the 
minimal energy configuration of repulsively interacting points on a 
2-manifold has become a standard problem of potential theory and has 
its paradigm in the classical Thomson problem on the sphere. The choice 
of the cubic potential is motivated here by the so called ``poppy seed bagel theorem''
\cite{HardinSaff:2005}, according to which the configuration of points 
that minimizes the Riesz energy $E=\sum_{i<j}1/|\bm{x}_{i}-\bm{x}_{j}|^{s}$
on a rectifiable manifold of Hausdorff dimension $d$ is uniformly 
distributed on the manifold for $s\ge d$. In the case of a torus of
revolution this implies that for small $s$ the points are mostly 
distributed on the exterior of the torus (the interior becomes 
completely empty in the limit $V\rightarrow\infty$). As $s$
is increased, however, the points cover a progressively larger
portion of the surface. The distribution becomes uniform for $s\ge 2$. 
On the other hand, since the number of local minima of the Riesz energy 
increases with $s$, it is practical to choose a value not 
much larger than two. The choice $s=3$
has the further advantage of modelling a real physical system of 
neutral colloidal particles assembled at an interface \cite{Pieranski:1980}
and is therefore suitable for direct comparison with experiments 
on colloidal suspensions.

To construct low energy configurations we adopt an carefully
designed hybrid optimization algorithm named Tapping (TA). Like other
hybrid algorithms, TA consists in a combination of fast local optimizations 
and global stochastic moves designed to release the system from the 
local minimum to which it is confined at the end of a local minimization step.
A more detailed description of our algorithm is reported in 
Appendix~\ref{app:3}.
We study four different aspect ratios: $r=3\,,4\,,6$ and $20$. For 
each aspect ratio we consider several different particle numbers 
up to $V=1000$ and each simulation is performed for $10^{5}$ to 
$10^{6}$ TA iterations. 

\begin{figure*}[h]
\centering
\includegraphics[width=1.8\columnwidth]{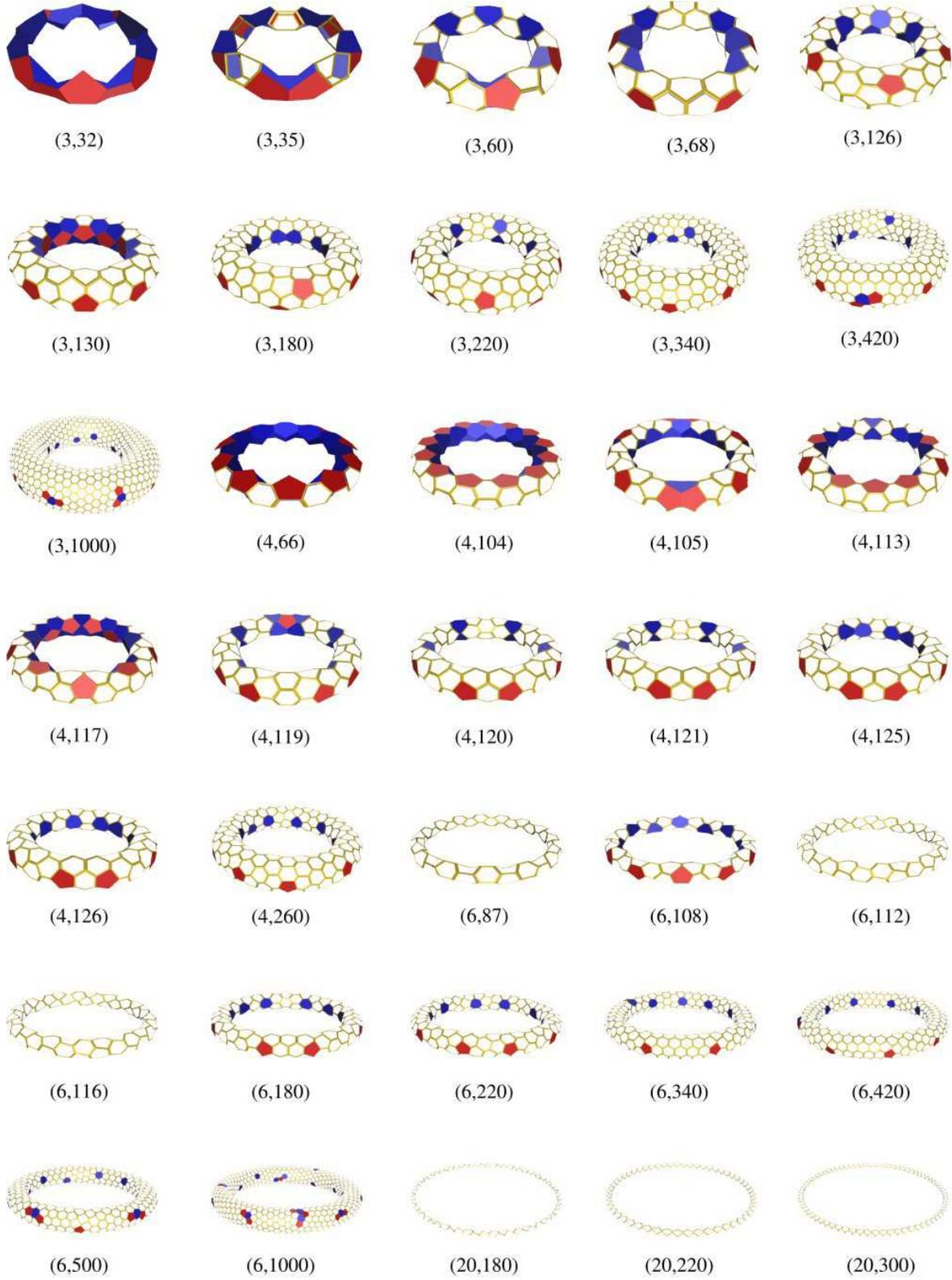}
\caption{\label{fig:numeric_tori}(Color online) Selected low energy configurations for
toroidal lattices of aspect ratio $r=3,\,4,\,6$ and $20$. Lattices are 
labeled by $(r,V)$, with $V$ the number of particles.}
\end{figure*}

The lowest energies found, as well as the number of defects in
the corresponding configuration, are reported in Table~\ref{tab:minima}
for a selected set of systems. The corresponding lattices are shown 
in Fig.~\ref{fig:numeric_tori} 

The lattices are best presented using a Voronoi construction 
corresponding to the dual lattice of the Delaunay triangulation. 
Here pentagonal faces are colored in red while heptagonal faces are 
colored in blue. The complete set of data produced in our simulations 
together with a collection of interactive 3D graphics for each low energy 
configuration found is available on-line~\cite{torusdatabase}.
For fewer than $V\sim 180$ particles the results of our 
numerical minimization are in good agreement with the continuum 
elastic theory. In particular for $180<V<500$ and $r=3,\,4$ and $6$, 
we always find minimal energy configurations consisting of ten $5-$fold 
disclinations on the outside of the torus and ten $7-$fold disclinations 
in the inside as predicted by the elastic theory in the regime
of $\epsilon_{c}/(AY)\sim 0$. For $r=20$ and $V>110$ we also find
the lowest energy configurations to be defect free. 

For small numbers of particles we don't expect the continuum 
approximation to accurately describe the lowest energy structure
of the toroidal clusters presented in Fig.~\ref{fig:numeric_tori}. 
Loosely speaking the limit of validity of the elastic theory can
be quantified by requiring the average lattice spacing 
$a=2\pi[R_{1}R_{2}/(\frac{\sqrt{3}}{2}V)]^{1/2}$ to be much smaller
than the radius $R_{2}$ of the torus. This condition requires $V$ to
be of order $500$ particles for a torus with aspect ratio $r=3$.
Remarkably, good agreement between the theory and simulations is 
found starting from much smaller values of $V$ and in some cases
(see the following discussion on the configuration with $r=3$ and
$V=130$), we already observe the onset of the ideal behavior predicted 
by theory for  $a\sim R_{2}$. The occurrence
of a ground state configuration with exact prismatic or antiprismatic
symmetry, in particular, is only possible when the number of particles
$V$ belongs to a specific sequence of ``magic numbers'' described in
Sec.~\ref{sec:3}. Nevertheless for $V$ outside such a sequence it is
still possible to observe in the ground state a predominant prismatic
or antiprismatic character depending on the aspect ratio. 

\begin{table}[h]
\begin{ruledtabular}
\begin{tabular}{c|cc}
$V$ & $r_{\min} \pm 0.05$ & $r_{\max} \pm 0.05$\\
\hline
16  & 1.0 & 1.6\\[2pt]
20  & 1.4 & 2.6\\[2pt]
24  & 1.8 & 3.4\\[2pt]
28  & 2.4 & 4.0\\[2pt]
32  & 2.9 & 4.6
\end{tabular}
\end{ruledtabular}
\caption{\label{tab:antiprism}Maximum and minimum aspect ratio for which
the toroidal antiprisms are a global minimum.}
\end{table}

Some configurations deserve special attention. For $V=16,\,20,\,24,\,28$
and $32$ and $r$ within a specific range (see Table \ref{tab:antiprism})
the global minima are represented by the second to sixth toroidal antiprims
discussed in Sec.~\ref{sec:3}. The drilled icosahedron, on the other hand,
would require the aspect ratio be less than one, as can be understood from 
Table \ref{tab:antiprism}, and is therefore never a minimum for $V=12$. We 
next describe the salient features of the four aspect ratios simulated.

$r=3$) The smallest minimal energy state with $D_{5h}$ symmetry 
is obtained for $V=35$. It features ten disclination pairs and belongs to the 
class of TP$5$a graphs. For $V=42$,  a TP$6$b lattice is obtained with no 
defects along the two equators. Two $6-$fold chiral configurations are 
obtained for $V=60$ and $126$. The global minimum obtained for $V=130$ 
displays a fascinating example of $5-$fold antiprismatic symmetry with 
the ten isolated negative disclinations in the interior of the torus 
replaced by a simplicial complex consisting of five triangles with a common 
$5-$fold apex and four $7-$fold coordinated vertices along the base.
A peculiar example is also represented by the minimum obtained for 
$V=180$. The lattice exhibits the typical pattern of a TP$n$ 
graph with $(3,1,4,7)-$type unit cell. The angular distance between 
neighboring disclinations is $\delta\phi\sim 2\pi/9$. Since
a prismatic graph cannot have an odd number of disclination pairs
the toroidal lattice is closed by a simplicial complex consisting of
two positive disclinations on the exterior of the torus at the opposite
sides of the external equator and two negative disclinations in the 
interior arranged similarly.  The total number of disclination
pairs is therefore ten. The typical pattern of the $5-$fold antiprismatic 
toroid can be found in all configurations with $V>200$. 
A single $5-7-5$ scar appears in the $r=3$ configurations at $V=420$,
while larger lattices (i.e $V=460$ and $500$) also feature $4-$fold
disclinations in the interior of the torus. It is not clear, however,
whether the presence of disclinations with topological charge $|q|>1$
is a genuine property of the ground state or rather an artifact due
to a misconvergence of our algorithm. 

\begingroup
\squeezetable
\begin{table}[h!]
\begin{ruledtabular}
\begin{tabular}{c|c|cccccc}
$r$ & $V$ & $V_{4}$ & $V_{5}$ & $V_{6}$ & $V_{7}$ & $V_{8}$ & Energy \\
\hline
\multirow{18}{*}{3} & 
  32 & 0 & 16 & 0 & 16 & 0 & 505.086593 \\
& 35 & 0 & 10 & 15 & 10 & 0 & 637.633663 \\
& 42 & 0 & 12 & 18 & 12 & 0 & 1020.466912 \\
& 120 & 0 & 14 & 92 & 14 & 0 & 14671.476332 \\
& 121 & 0 & 20 & 81 & 20 & 0 & 14981.224344 \\
& 125 & 0 & 19 & 87 & 19 & 0 & 16255.583992 \\
& 126 & 0 & 12 & 102 & 12 & 0 & 16586.793347 \\
& 130 & 0 & 20 & 90 & 20 & 0 & 17930.955152 \\
& 180 & 0 & 10 & 160 & 10 & 0 & 40623.325218 \\
& 220 & 0 & 10 & 200 & 10 & 0 & 67176.585493 \\
& 260 & 2 & 16 & 222 & 20 & 0 & 102100.926892 \\
& 300 & 1 & 10 & 277 & 12 & 0 & 146139.605664 \\
& 340 & 0 & 10 & 320 & 10 & 0 & 199812.441922 \\
& 420 & 0 & 11 & 398 & 11 & 0 & 339147.966681 \\
& 460 & 2 & 14 & 426 & 18 & 0 & 425754.968401 \\
& 500 & 2 & 16 & 462 & 20 & 0 & 524508.172150 \\
& 1000 & 1 & 17 & 963 & 19 & 0 & 2965940.674307 \\
\hline
\multirow{18}{*}{4} &
  66 & 0 & 22 & 22 & 22 & 0 & 4905.964854 \\
& 104 & 0 & 26 & 52 & 26 & 0 & 15598.534409 \\
& 105 & 0 & 15 & 75 & 15 & 0 & 15984.990289 \\
& 113 & 0 & 19 & 75 & 19 & 0 & 19237.981548 \\
& 117 & 0 & 22 & 73 & 22 & 0 & 21007.172188 \\
& 119 & 0 & 12 & 95 & 12 & 0 & 21914.283713 \\
& 120 & 0 & 10 & 100 & 10 & 0 & 22371.402771 \\
& 121 & 0 & 12 & 97 & 12 & 0 & 22859.735385 \\
& 125 & 0 & 10 & 105 & 10 & 0 & 24816.591295 \\
& 126 & 0 & 10 & 106 & 10 & 0 & 25311.298095 \\
& 180 & 0 & 10 & 160 & 10 & 0 & 62142.129092 \\
& 220 & 0 & 10 & 200 & 10 & 0 & 102919.127703 \\
& 260 & 0 & 10 & 240 & 10 & 0 & 156499.285669 \\
& 300 & 0 & 10 & 280 & 10 & 0 & 223997.341297 \\
& 340 & 0 & 10 & 320 & 10 & 0 & 306568.539431 \\
& 420 & 0 & 13 & 394 & 13 & 0 & 520431.653442 \\
& 460 & 0 & 11 & 438 & 11 & 0 & 653485.181907 \\
& 500 & 0 & 14 & 472 & 14 & 0 & 805206.972227 \\
\hline
\multirow{11}{*}{6} &
   87 & 0 & 0 & 87 & 0 & 0 &  17765.124942 \\
& 108 & 0 & 12 & 84 & 12 & 0 & 30894.374674 \\
& 112 & 0 & 0 & 112 & 0 & 0 & 33902.717714 \\
& 115 & 0 & 0 & 115 & 0 & 0 & 36254.709031 \\
& 116 & 0 & 0 & 116 & 0 & 0 & 37074.949162 \\
& 180 & 0 & 10 & 160 & 10 & 0 & 112810.451302 \\
& 220 & 0 & 10 & 200 & 10 & 0 & 187146.462245 \\
& 260 & 0 & 10 & 240 & 10 & 0 & 284907.016076\\
& 340 & 0 & 10 & 320 & 10 & 0 & 559161.546358 \\
& 420 & 0 & 10 & 400 & 10 & 0 & 950488.931696 \\
& 500 & 0 & 13 & 474 & 13 & 0 & 1471923.063515 \\
& 1000 & 0 & 30 & 940 & 30 & 0 & 8351619.696538 \\
\hline 
\multirow{6}{*}{20} &
  160 & 0 & 0 & 160 & 0 & 0 & 463967.242489 \\
& 170 & 0 & 0 & 160 & 0 & 0 & 543799.839326 \\
& 180 & 0 & 0 & 180 & 0 & 0 & 631751.371902 \\
& 220 & 0 & 0 & 220 & 0 & 0 & 1065625.748639 \\
& 260 & 0 & 0 & 260 & 0 & 0 & 1636942.532923 \\
& 300 & 0 & 0 & 300 & 0 & 0 & 2370110.403872 \\
\end{tabular}
\end{ruledtabular}
\caption{\label{tab:minima}Low energy configuration for a selected 
number of toroidal lattices with aspect ratios $r=3,\,4,\,6$ and $20$.
For each aspect ratio the table displays the number of particles $V$,
the lowest energy found and the number of $k-$fold vertices $V_{k}$
with $k=4$--$8$.
}
\end{table}
\endgroup

$r=4$) An interesting feature is observed at $V=42$. As in the case 
of $r=3$ we also find a minimum with $D_{6h}$ symmetry group, but unlike 
the latter configuration, it belongs to the TP$6$a class and has 
$5-$fold disclinations along the external equator. A TP$7$b configuration 
is obtained again for $V=49$. For $V=66$ and $V=104$ the global minimum 
is achieved by two spectacular antiprismatic configurations with $D_{11d}$
and $D_{13d}$ symmetry group respectively. These toroids can be obtained
from the toroidal antiprisms discussed in Sec.~\ref{sec:3} by 
splitting~\footnote{\emph{Vertex splitting} is a standard operation to
generate larger triangulations from an irreducible one. It consists in
dividing an existing vertex $v$ in two such that the total number of 
vertices is increased by one. The two newly created vertices have
coordination number $c-1$, with $c$ is the coordination number $v$,
while two of the $c$ neighbors of $v$ gain a bond.} one or more times 
the initial set of $5-$fold vertices. Thus starting from a $11-$fold 
toroidal antiprism with $V=4\times 11=44$ vertices and splitting all 
$22$ $5-$fold vertices one obtains $V=44+22=66$ vertices. Splitting 
twice all $26$ $5-$fold vertices of a $13-$fold toroidal antiprism 
with $V=4\times 13=52$, on the other hand, we have $V=52+2\times 26=104$.
For $V=120$ the global minimum is represented by a fascinating 
lattice of TP$5$a type. Lattices with $V=121\,,125$ and $126$ resemble 
very closely the structure of a TP$5$ graph while for $V=260$ the 
lattice has a more antiprismatic character with ten defect pairs. 

$r=6$) Three defect-free configurations are found at $V=87,\,112$ and 
$116$. The case $V=87$ is a particular example of a defect-free lattice
that cannot be obtained form the planar construction reviewed in 
Sec.~\ref{sec:3}. It consists of $29$ octahedra connected in the form 
of a chain. Since each octahedron is attached to other two, it 
contributes with six faces to the total face count. Thus 
$F=6\times 29=174$ and $V=174/2=87$. For $180\le V < 460$ we always 
find configurations with ten disclination pairs as expected from continuum 
elasticity. For $V>460$ the regime of coexistence between isolated 
disclinations and scars described in Sec. \ref{sec:scars} is observed.
The delocalization of isolated disclinations into scars, however, doesn't take 
place at each defective site simultaneously and the regime of coexistence 
between positively charged scars and isolated $7-$fold disclinations is preceded
by a phase with isolated $5-$ and $7-$fold disclinations and scars.
For tori with aspect ratio as large as $r=20$ we find defect-free
ground states every time it is possible to construct a purely $6-$valent
toroidal graph with the same number of vertices $V$.

\section{\label{sec:6}The Fat Torus Limit}

We have seen that disclination defects, forbidden in the lowest energy state of a 
planar crystal, may be energetically favored on a substrate of  non-vanishing 
Gaussian curvature. It is therefore natural to ask whether large curvature can
completely destroy crystalline order by driving the proliferation of a sufficiently
high density of defects. The resulting state would be amorphous. The problem of 
generating amorphous structures by tiling a two-dimensional curved space with 
identical rigid subunits has drawn attention over the years, particularly through
the connection to the structure of such disordered materials as supercooled liquids
and metallic glasses. Since the work of Frank \cite{Frank:1952} the notion of
geometrical frustration arises frequently in investigations of supercooled liquids 
and the glass transition. A paradigmatic example is represented by the icosahedral 
order in metallic liquids and glasses which, although locally favored, cannot 
propagate throughout all of three-dimensional Euclidean space. A two-dimensional 
analog, consisting of a liquid of monodisperse hard disks in a 2-manifold of 
constant negative Gaussian curvature (the hyperbolic plane) was first proposed by 
Nelson and coworkers in 1983 \cite{Nelson:1983}. In such a system the impossibility
of covering the entire manifold with a 6-fold coordinated array of disks mimics 
many aspects of the geometrical frustration of icosahedral order in three 
dimensions. In all these models of geometrical frustration, however, the origin 
of the disorder is primarily due to the short-range nature of the potential 
between the subunits. In a more realistic setting, part of the frustration is 
relieved by the fact that hexagonal unit cells can compress in order to match the 
underlying geometry. 
  
The embedding of a triangular lattice on an axisymmetric torus, provides a 
particularly suitable playground to study curvature-driven disorder. When 
$r\rightarrow 1$ the Gaussian curvature on the inside of the torus grows
like $1/(r-1)$ and diverges on the internal equator at $\psi=\pi$. We thus expect 
a high density of defects in the vicinity of the curvature singularity and a 
resultant loss of the local $6-$fold bond orientational order. In this regime 
the system will have crystalline regions on the outside of the torus and 
amorphous regions near the curvature singularity.  

\begin{figure}[t]
\centering
\includegraphics[width=0.8\columnwidth]{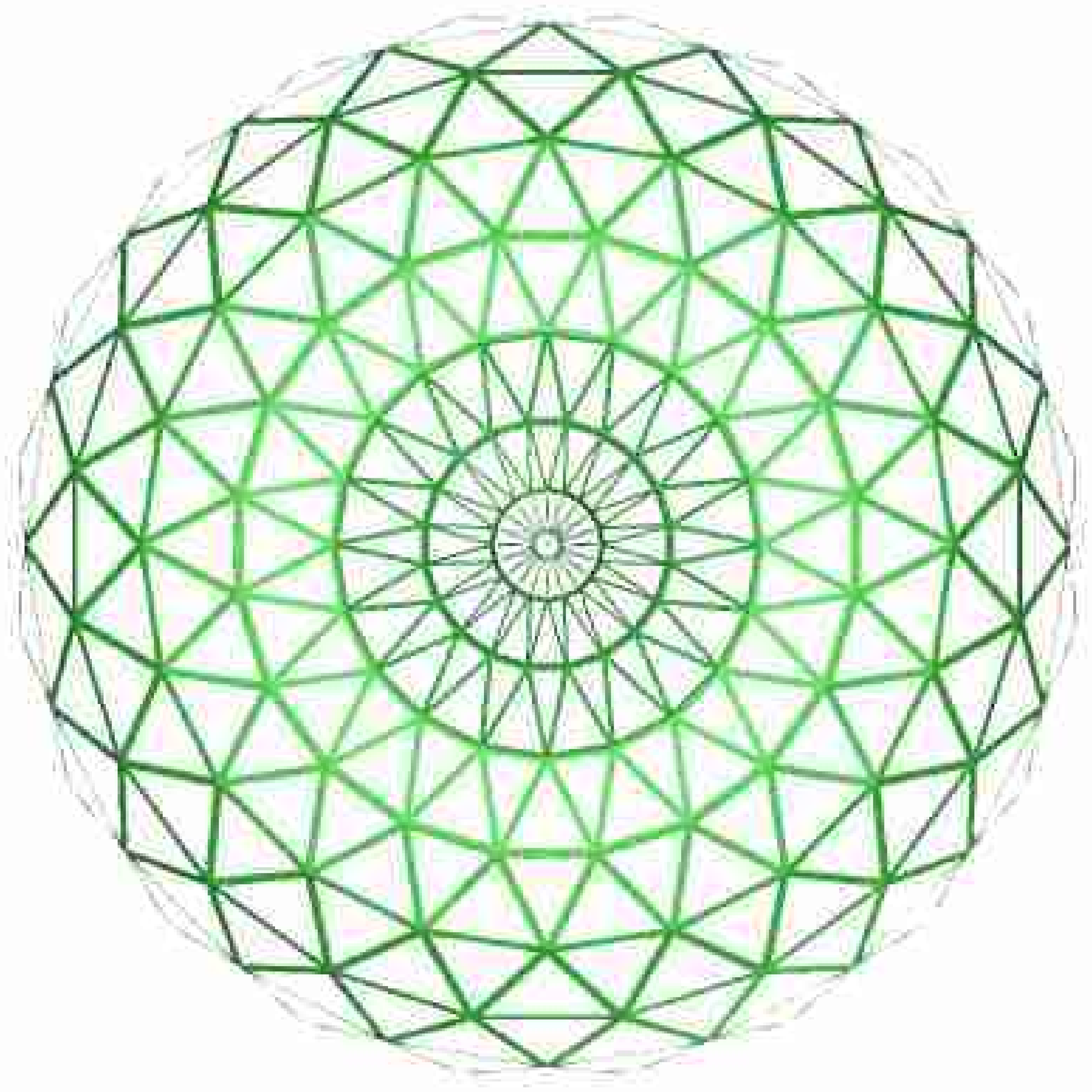}
\caption{\label{fig:fat_torus}(Color online) Top view of a defect free triangulation 
of a fat torus with $(n,m,l)=(10,10,20)$ and $V=400$. The corresponding
elastic energy becomes very high in the interior of the torus where the
triangles are more compressed to match the reduction of surface area.}
\end{figure}

\begin{figure}
\centering
\includegraphics[width=0.6\columnwidth]{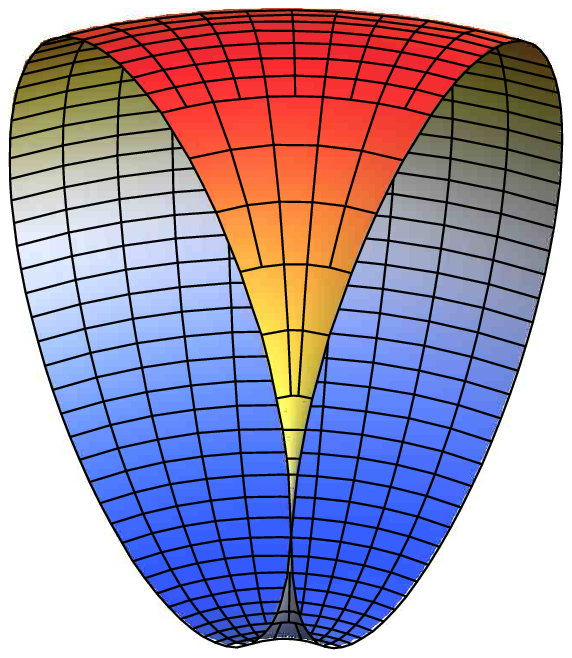}
\caption{\label{fig:dislocation}(Color online) A schematic example of a dislocation 
pile-up on a square lattice resulting from the shrinking of the area on a regular 
wedge of a fat torus.}
\end{figure}

In this section we substantiate this claim analytically based on the elastic 
theory of continuous distributions of edge dislocations on a ``fat'' torus.  
Our argument is based on the following construction. As a consequence of the 
curvature singularity the surface area of an arbitrary wedge of angular width 
$\Delta\phi$ becomes smaller and smaller as the sectional angle $\psi$ increases 
and vanishes at $\psi=\pi$. If a defect-free lattice is embedded on such a wedge, 
Bragg rows will become closer and closer as the singularity is approached with a 
consequent rise in the elastic energy (see Fig.~\ref{fig:fat_torus}). An intuitive 
way to reduce the distortion of the lattice is to recursively remove Bragg rows as one 
approaches the point $\psi=\pi$ (see Fig. \ref{fig:dislocation}). This is equivalent to 
introducing a growing density of edge dislocations. This dislocation ``cloud'' will 
ultimately disorder the system by destroying the local $6-$fold bond orientational order. 
One might therefore view the curvature as playing the role of a local effective temperature
which can drive ``melting'' by liberating disclinations and dislocations.
In two-dimensional non-Euclidean crystals at $T=0$, however, the mechanism for 
dislocation proliferation is fundamentally different from the usual thermal 
melting. While the latter is governed by an entropy gain due to unbinding of 
dislocation pairs, the amorphization at $T=0$ is due to the adjustment of the 
lattice to the geometry of the embedding manifold via the proliferation
of defects and the consequent release of elastic stress. A similar 
phenomenon occurs in the disorder-driven amorphization of vortex 
lattices in type-II and high-$T_{c}$ superconductors \cite{KierfeldVinokur:2000}.

Since the shrinking area per plaquette on the inside of the torus necessitates
a high density of dislocations we may approximate the dislocation cloud in this
region by a continuous distribution of Burgers vector density $\bm{b}$.
Minimizing the elastic energy with respect to $\bm{b}$ yields a variational 
equation from which the optimal dislocation density can be calculated as a 
function of the ratio $\epsilon_{d}/(YR^{2})$ between the dislocation core energy 
$\epsilon_{d}$ and elastic energy scale $YR^{2}$ with $R=R_{1}=R_{2}$. 

As a starting point, we calculate the Green function $G_{L}(\bm{x},\bm{y})$ in the
fat torus limit $r\rightarrow 1$ (i.e.  $\kappa\rightarrow\infty$ and 
$\omega\rightarrow 0$, see Eq.~\eqref{eq:kappa-omega}). The conformal angle $\xi$
in this limit is:
\[
\lim_{r\rightarrow 1} \xi = \tan\frac{\psi}{2}\,,
\]
and to leading order of $\kappa$ we have:
\begin{gather*}
\frac{\kappa}{16\pi^{2}}\left(\psi-\frac{2}{\kappa}\xi\right)^{2}
\rightarrow \frac{\kappa}{16\pi^{2}}\,\psi^{2}-\frac{\psi}{4\pi}\tan\frac{\psi}{2}\,,\\[7pt]
\frac{\kappa}{4\pi^{2}}\Real\{\Li(\alpha e^{i\psi})\}
\rightarrow \frac{\kappa}{16\pi^{2}}\,\psi^{2}+\frac{1}{2\pi^{2}}\log\left(\cos\frac{\psi}{2}\right)\,.
\end{gather*}
To handle the limit of the Jacobi theta function we can take
$u=\Delta z/\kappa$, $q=e^{i\pi\tau}=e^{-\frac{2\pi}{\kappa}}$ and 
calculate the limit $q\rightarrow 1$. This can be done by using 
the modular transformation properties of Jacobi functions\cite{Theta}:
\begin{equation}\label{eq:modular}
\vartheta_{1}\left(\tfrac{u}{\tau}|-\tfrac{1}{\tau}\right) = -i(-i\tau)^{\frac{1}{2}}e^{\frac{iu^{2}}{\pi\tau}}\vartheta_{1}(u|\tau)\,.
\end{equation}
Thus $\tau' = -1/\tau = i\kappa/2$, $u' = u/\tau = \Delta z/(2i)$ 
and $q'=e^{i\tau'}=e^{-\frac{\pi\kappa}{2}}$, where:
\[
\lim_{q\rightarrow 1}\vartheta_{1}(u,q) 
= \lim_{q'\rightarrow 0} i\left(\frac{i}{\tau'}\right)^{-\frac{1}{2}} e^{\frac{iu'^{2}}{\pi\tau'}}\vartheta_{1}(u',q')\,.
\]
This is easily evaluated by means of the expansion:
\[
\vartheta_{1}(u,q) = 2q^{\frac{1}{4}}\sin u + o\left(q^{\frac{9}{4}}\right)\,.
\]
Taking the logarithm and neglecting irrelevant constant terms, we obtain:
\[
\log \left|\vartheta_{1}\left(\frac{z-z'}{\kappa}\Bigg|\frac{2i}{\kappa}\right)\right| \sim \log\left|\sinh\left(\frac{z-z'}{2}\right)\right|\,, 
\]
which finally leads to:
\begin{multline}
G_{L}(\psi,\phi,\psi',\phi')
\sim -\frac{\psi'}{4\pi^{2}}\,\tan\frac{\psi'}{2}\\
+\frac{1}{2\pi}\log\left|\sinh\left(\frac{z-z'}{2}\right)\right|
\end{multline}
with $z=\tan(\psi/2)+i\phi$. With the Green function in hand, we can
calculate the effect of the curvature singularity at $\psi=\pi$ 
on the distribution of defects. Let $\bm{b}$ be the Burgers vector density 
of the dislocation cloud. Hereafter we work in a local frame, so that:
\begin{equation}\label{eq:burger}
\bm{b}=b^{\psi}\bm{g}_{\psi}+b^{\phi}\bm{g}_{\phi}\,,
\end{equation}
with $\bm{g}_{i}=\partial_{i}\bm{R}$ a basis vector in the tangent 
plane of the torus whose points are specified by the three-dimensional 
Euclidean vector $\bm{R}$. The quantity $\bm{b}$ has to be such that:
\[
\int_{D} d^{2}x\,\bm{b}(x)=\bm{b}_{D}\,,
\]
with $\bm{b}_{D}$ the total Burger's vector in a generic domain $D$. Because
on a closed manifold dislocation lines cannot terminate on the boundary, extending
the integration to the whole torus we have:
\begin{equation}\label{eq:burger_neutrality}
\int d^{2}x\,\bm{b}(\bm{x})=0\,.
\end{equation}
Since the basis vectors $\bm{g}_{i}$ in Eq.~\eqref{eq:burger} have the dimension 
of length, contravariant coordinates $b^{i}$ have dimensions of an inverse area. 
Assuming all defects to be paired in the form of dislocations (i.e. $q_{i}=0$ 
everywhere), the total energy of the crystal reads:
\begin{equation}\label{eq:cloud_energy}
F= \frac{1}{2Y} \int d^{2}x\, \Gamma^{2}(\bm{x}) + \epsilon_{d}\int d^{2}x\,|\bm{b}(\bm{x})|^{2}\,,
\end{equation}
where $\epsilon_{d}$ is the dislocation core energy and
\[
|\bm{b}|^{2}=g_{ij}b^{i}b^{j}=g_{\psi\psi}(b^{\psi})^{2}+g_{\phi\phi}(b^{\phi})^{2}\,.
\] 
The function $\Gamma(\bm{x})$ encoding the elastic stress due to the curvature 
and the screening contribution of the dislocation cloud obeys
\begin{equation}\label{eq:cloud_poisson}
\frac{1}{Y}\Delta_{g}\Gamma(\bm{x})=\epsilon_{k}^{i}\nabla_{i}b^{k}(\bm{x})-K(\bm{x})\,,
\end{equation}
where $\epsilon_{k}^{i}$ is the Levi-Civita antisymmetric tensor on the torus:
\[
\epsilon_{\psi\phi}=-\epsilon_{\phi\psi}=\sqrt{g}\,,
\qquad\qquad
\epsilon_{i}^{j} = g_{ik}\epsilon^{jk}\,.
\]
The stress function $\Gamma(\bm{x})$ can be expressed in the form 
$\Gamma(\bm{x}) = \Gamma_{d}(\psi,\phi)-\Gamma_{s}(\psi)$ with
\begin{subequations}
\begin{gather}
\frac{\Gamma_{s}(\psi)}{Y} = \log\left[\frac{1}{2(1+\cos\psi)}\right]+1\,,\label{eq:fat_gamma_screening}\\[7pt]
\frac{\Gamma_{s}(\psi,\phi)}{Y} = \int d^{2}y\,\epsilon_{k}^{i}\nabla_{i}b^{k}(\bm{y})\,G_{L}(\bm{x},\bm{y})\,.\label{eq:fat_gamma_defects}
\end{gather}
\end{subequations}
Taking advantage of the closeness of the torus we can integrate 
Eq.~\eqref{eq:fat_gamma_defects} by parts so that:
\begin{equation}\label{eq:cloud_gamma1}
\frac{\Gamma_{d}(\psi,\phi)}{Y} = -\int d^{2}y\,\epsilon_{k}^{i}b^{k}(\bm{y})\partial_{i}\,G_{L}(\bm{x},\bm{y})\,.
\end{equation}
Now we want reduce the integral term in Eq. \eqref{eq:cloud_gamma1} 
to a more friendly functional of $\bm{b}$, suitable for a variational 
approach. Given the azimuthal symmetry we assume that all dislocations
are aligned along $\bm{b}=b^{\phi}\bm{g}_{\phi}$. Even though 
not necessarily true, we argue this to be a reasonable work hypothesis 
as well as a solid starting point to capture the essential physics of 
the fat limit. In this case $\Gamma_{d}(\psi,\phi)=\Gamma_{d}(\psi)$ can 
be recast in the form
\begin{multline}\label{eq:cloud_gamma2}
\frac{\Gamma_{d}(\psi)}{Y}
= \frac{1}{2\pi}\int_{-\pi}^{\pi} d\psi'\sqrt{g}\,b^{\phi}(\psi')\\\left[\psi'+\sin\psi'+\pi\sgn(\psi-\psi')\right]\,.
\end{multline}
Substituting Eq.~\eqref{eq:cloud_gamma2} and \eqref{eq:fat_gamma_screening} 
in Eq.~\eqref{eq:cloud_energy} and minimizing with respect to $b^{\phi}$ we 
can now write the variational equation:
\begin{gather}
4\epsilon_{d}R^{2}(1+\cos\psi)^{2}b^{\phi}(\psi)
+\int_{-\pi}^{\pi} d\psi\,\sqrt{g}\,\Gamma_{d}(\psi')\sgn(\psi-\psi')\notag\\
=\int_{-\pi}^{\pi} d\psi\,\sqrt{g}\,\Gamma_{s}(\psi')\sgn(\psi-\psi')\,.\label{eq:fredholm1}
\end{gather}
By inverting the order of integration in the integral on the right hand 
side, Eq~\eqref{eq:fredholm1} can be expressed in the form of a Fredholm 
equation of the second kind:
\begin{equation}\label{eq:fredholm2}
\lambda B(\psi) - \int_{-\pi}^{\pi} d\psi'\,B(\psi')\mathcal{K}(\psi,\psi')=f(\psi)\,,
\end{equation}
where $\lambda=\epsilon_{d}/(YR^{2})$, $B(\psi)=R^{2}(1+\cos\psi)^{2}b^{\phi}(\psi)$ 
and the kernel $\mathcal{K}(\psi,\psi')$ is given by:
\begin{multline}
\mathcal{K}(\psi,\psi')=
\frac{1}{4(1+\cos\psi')}\Bigg\{\pi^{-1}(\psi'+\sin\psi')(\psi+\sin\psi)\\
+|\psi-\psi'|+2\cos\frac{\psi+\psi'}{2}\sin\frac{|\psi-\psi'|}{2}\Biggr\}\,.
\end{multline}
The function $f(\psi)$ on the right hand side of Eq.~\eqref{eq:fredholm2} 
is given by:
\begin{gather}\label{eq:fredholm_known1}
f(\psi)
= -\frac{1}{4}\int_{-\psi}^{\psi} d\psi'\,(1+\cos\psi')\,\Gamma_{s}(\psi')\notag\\
= \frac{1}{2}\bigl\{[\log 2(1+\cos\psi)-2]\sin\psi-2\Cl(\psi+\pi)\bigr\}
\end{gather}
where $\Cl$ is the Clausen function (see Ref. \onlinecite{AbramowitzStegun}, 
pp. 1005-1006) defined as:
\[
\Cl(x)
=-\int_{0}^{x} dx\, \log\left(2\sin\frac{t}{2}\right)
= \sum_{k=1}^{\infty}\frac{\sin kx}{k^{2}}\,.
\]
\begin{figure}[t]
\centering
\includegraphics[width=1.\columnwidth]{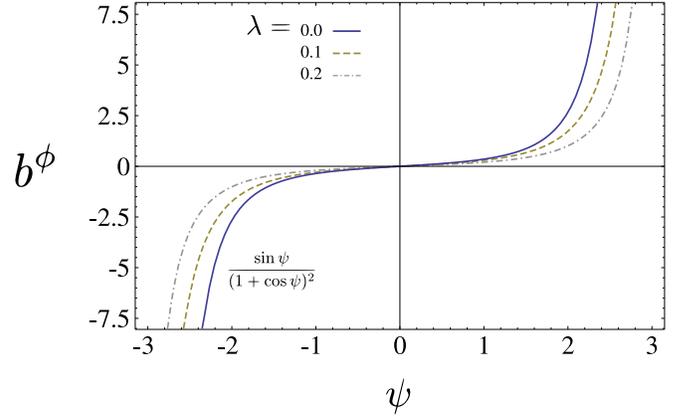}
\caption{\label{fig:burgers}(Color online) The Burgers vector component $b(\psi)$ 
for different choices of $\lambda$.}
\end{figure}

As previously noted the dislocation core energy is $\epsilon_{d}$ is much
smaller than the elastic energy scale $YR^{2}$. Eq~\eqref{eq:fredholm2} is 
then suitable to be solved in powers of the dimensionless number $\lambda$:
\[
B(\psi) = B_{0}(\psi)+\lambda B_{1}(\psi)+\lambda^{2} B_{2}(\psi)+\cdots
\]
The corrections to the zero-order term $B_{0}(\psi)$ can be calculated 
recursively by solving a set of Fredholm equations of the first kind:
\[
B_{k-1}(\psi) = \int_{-\pi}^{\pi} d\psi'\,B_{k}(\psi)\mathcal{K}(\psi,\psi')\qquad k\ge 1\,.
\]
The function $B_{0}(\psi)$ associated with the Burgers vector density 
of the dislocation cloud in the limit $\lambda\rightarrow 0$, on 
the other hand, can be calculated directly from Eq.~\eqref{eq:cloud_poisson} 
by setting the effective topological charge density on the right 
hand side to zero:
\begin{equation}\label{eq:zero_core1}
\epsilon_{k}^{i}\nabla_{i}b^{k}(\bm{x})-K(\bm{x}) = 0\,.
\end{equation}
For a torus of revolution the only nonzero Christoffel symbols are
\begin{gather*}
\Gamma_{\phi\psi}^{\phi} = \Gamma_{\psi\phi}^{\phi} = -\frac{R_{2}\sin\psi}{R_{1}+R_{2}\cos\psi}\,,\\[7pt]
\Gamma_{\phi\phi}^{\psi} = R_{2}^{-1}\sin\psi(R_{1}+R_{2}\cos\psi)\,.
\end{gather*}
Since $b^{\psi}=0$ by assumption, the first term in Eq.~\eqref{eq:zero_core1} can be expressed as:
\begin{align*}
\epsilon_{k}^{i}\nabla_{i}b^{k}
&=\epsilon_{\phi}^{\psi}(\partial_{\psi}b^{\phi}
+ \Gamma_{\psi\phi}^{\phi}b^{\phi})+\epsilon_{\psi}^{\phi}\Gamma_{\phi\phi}^{\psi}b^{\phi}\\[7pt]
&=(1+\cos\psi)\partial_{\psi}b^{\phi}-2\sin\psi\,b^{\phi}\,,
\end{align*}
and Eq.~\eqref{eq:zero_core1} becomes an ordinary differential equation
\begin{equation}\label{eq:zero-core2} 
\partial_{\psi}b^{\phi}-\frac{2\sin\psi}{1+\cos\psi}\,b^{\phi}=\frac{\cos\psi}{R^{2}(1+\cos\psi)^{2}}\,,
\end{equation}
whose solution is given by
\begin{equation}\label{eq:zero-order}
b^{\phi} = \frac{\sin\psi}{R^{2}(1+\cos\psi)^{2}}\,,
\end{equation}
so that $B_{0}(\psi)=\sin\psi$. The Burgers vector density $b^{\phi}$ obtained 
by a numerical solution of Eq.~\eqref{eq:fredholm2} is shown in Fig.~\ref{fig:burgers} 
for different values of $\lambda$. The Burgers vector density is measured in units
 of $R^{-2}.$ The function $b^{\phi}$ has cubic singularities at $\psi=\pm\pi$ and 
is approximately zero on the outside of the torus. The solid blue curve in 
Fig.~\ref{fig:burgers} represents the zeroth order solution of Eq.\eqref{eq:zero-order}.

\begin{figure}[h!]
\centering
\includegraphics[width=1.\columnwidth]{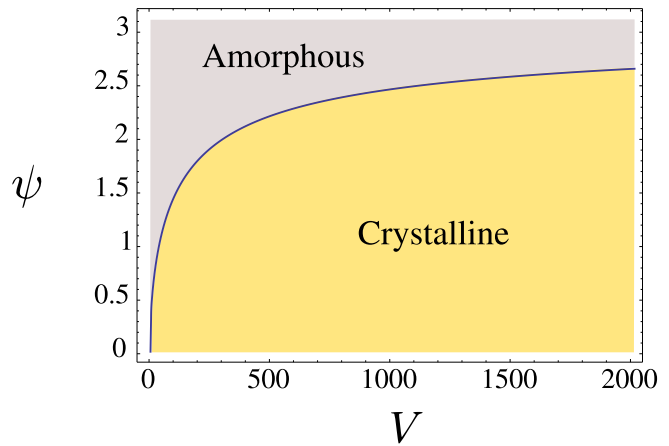}
\caption{\label{fig:melting-diagram}(Color online) Phase diagram for curvature driven 
amorphization. The inside of the torus contains an amorphous region whose angular 
size decreases with the number of vertices $V$ as a consequence of the reduction 
of the lattice spacing.}
\end{figure}

Now, in the theory of dislocation mediated melting a system at the solid liquid 
phase boundary is described as a crystalline solid saturated with dislocations.
In three-dimensions, in particular, there is a strong experimental evidence of
the existence of a critical dislocation density at the melting point 
$\rho(T_{m})\approx 0.6b^{-2}$ where $b$ is the length of the length of the 
smallest perfect-dislocation Burgers vector \cite{BurakovskyEtAl:2000}. Several
theoretical works have motivated this evidence both for three-dimensional solids
and vortex lattices in super conductors \cite{KierfeldVinokur:2000}. On the other
hand, given the existence of such a critical density, its value can be empirically
used to determine whether a system is in a solid or liquid-like phase in the same
spirit as the Lindemann criterion. With this goal in mind we can calculate 
the dislocation density by requiring $|\bm{b}|=\rho a$ with $\rho$ the density 
of single lattice spacing dislocations. This yields:
\begin{equation}\label{eq:meltin-criterion}
\rho(\psi,V) a^{2} = 2\pi \left(\frac{\sqrt{3}}{2}\,V\right)^{-\frac{1}{2}}\left|\tan\frac{\psi}{2}\right|+o(\lambda)
\end{equation}
Solving $\rho(\psi,V)a^{2}=0.6$ as a function of $\psi$ and $V$ we obtain the 
diagram of Fig.~\ref{fig:melting-diagram}. As expected the inside of the torus
contains an amorphous region whose angular size decreases with the number of 
vertices $V$ as a consequence of the reduction of the lattice spacing.

\section{\label{sec:7}Discussion and Conclusions}

In this article we reported a comprehensive analysis of the ground state 
properties of torodial crystals. Using the elastic theory of defects on
curved substrates we identified the ground state structures of an arbitrary
crystalline torus as a function of the aspect ratio $r$ and the ratio 
$\epsilon_{c}/(AY)$ of the defect core energy to the elastic energy scale.
We showed that for a large range of aspect ratios and core energies the 
minimal energy structure of a toroidal crystal has ten disclinations 
pairs and symmetry group $D_{5h}$, as first conjectured by Lambin 
\emph{et al} more than ten years ago~\cite{LambinEtAl:1995}. For large system
sizes we proved isolated disclinations are unstable to grain-boundary 
scars consisting of chains of tightly bound $5-7$ pairs radiating
form an unpaired disclination. On a torus, where the Gaussian curvature
on the inside is always larger in magnitude than that on the outside, the 
occurrence of scars is marked by a state featuring isolated 
$7-$fold disclinations in the inside of the torus together with scars on the 
outside. For a torus of aspect ratio close to one we showed how a
diverging Gaussian curvature on the internal equator is responsible
for the remarkable occurrence of a curvature-driven transition of the
system to a disordered, liquid-like, state. The predictions of our elastic 
theory were compared with the results of a numerical study of 
a system of point-like particles constrained on the surface of a torus 
and interacting via a short range potential, with  good agreement. 
From a purely geometrical point of view we have introduced a number
of novel toroidal polyhedra as well as a new construction and 
classification scheme for certain types of prismatic tori.

\acknowledgments

This project was supported by NSF grants DMR-0305407 and DMR-0705105
and by an allocation through the TeraGrid Advanced Support Program. 
LG is supported on a Graduate Fellowship by the Syracuse Biomaterials
Institute. We acknowledge Amand Lucas and Antonio Fonseca for sharing
with us the beautiful photos presented in Sec.\ref{sec:3}.
We also would like to thank Eric J. West and M. Cristina 
Marchetti for inspiring discussions during the development of Sec.
\ref{sec:6}. LG is very grateful to Dario Giuliani who made 
possible the realization of the \emph{Toroidal Lattices
Database}~\cite{torusdatabase}, the use of which has been crucial 
for obtaining many of the results and the ideas
presented in this paper.

\appendix

\section{\label{app:1}The Green Function on the Torus}

The modified Laplacian Green function on a periodic rectangle of edges 
$p_{1}$ and $p_{2}$ can be conveniently calculated in the form:
\begin{equation}
G_{0}(\bm{x},\bm{y}) = \sum_{\lambda \ne 0} \frac{u_{\lambda}(\bm{x})\overline{u}_{\lambda}(\bm{y})}{\lambda}\,,
\end{equation}
where $u_{\lambda}$ is the eigenfunction of the Laplace operator with 
periodic boundary conditions:
\begin{equation}
\Delta u_{\lambda}(\bm{x}) = \lambda u_{\lambda}(\bm{x})\,,
\end{equation}
such that:
\[
\left\{
\begin{array}{c}
u_{\lambda}(0,\eta) = u_{\lambda}(p_{1},\eta) \\[7pt]
u_{\lambda}(\xi,0) = u_{\lambda}(\xi,p_{2})
\end{array}
\right.\,.
\]
In Cartesian coordinates the eigenfunctions are simple plane waves of the form:
\begin{equation}
u_{\lambda}(\xi,\eta) = \frac{e^{i(\lambda_{n}\xi+\mu_{m}\eta)}}{\sqrt{p_{1}p_{2}}}\,,
\end{equation}
where $\lambda_{n}$ and $\mu_{m}$ are given by:
\[
\lambda_{n} = \frac{2\pi n}{p_{1}} \qquad
\mu_{m} = \frac{2\pi m}{p_{2}} \qquad
n,\,m = 0,\,\pm 1,\,\pm 2\ldots
\]
and the eigenvalue $\lambda$ is given by:
\begin{equation}
\lambda = - \lambda_{n}^{2} - \mu_{m}^{2}\,.
\end{equation}
Calling for simplicity $\bm{x}=(x,y)$ and $\bm{y}=(\xi,\eta)$, the function 
$G_{0}$ is given by:
\begin{equation}\label{eq:g0_1}
G_{0}(\bm{x},\bm{y}) 
= -\frac{1}{p_{1}p_{2}}\sum_{(n,m)\ne(0,0)} \frac{e^{i\lambda_{n}(x-\xi)}e^{i\mu_{m}(y-\eta)}}{\lambda_{n}^{2}+\mu_{m}^{2}}\,.
\end{equation}
Although \eqref{eq:g0_1} is simple, it is very useful to rewrite it in terms
of elliptic functions. Noting that the odd terms in \eqref{eq:g0_1} cancel
we have:
\begin{widetext}
\begin{align}\label{eq:g0_2}
G_{0}(\bm{x},\bm{y})
&=-\frac{1}{p_{1}p_{2}} \sum_{(n,m)\ne(0,0)}\frac{\cos\lambda_{n}(x-\xi)\cos\mu_{m}(y-\eta)}{\lambda_{n}^{2}+\mu_{m}^{2}}\notag\\[7pt]
&=-\frac{2}{p_{1}p_{2}} \left[
  \sum_{m=1}^{\infty}\frac{\cos\frac{2\pi m}{p_{2}}(y-\eta)}{\left(\frac{2\pi m}{p_{2}}\right)^{2}}+
  \sum_{n=1}^{\infty}\sum_{m=-\infty}^{\infty}\frac{\cos\frac{2\pi n}{p_{1}}(x-\xi)\cos\frac{2\pi m}{p_{2}}(y-\eta)}
   {\left(\frac{2\pi n}{p_{1}}\right)^{2}+\left(\frac{2\pi m}{p_{2}}\right)^{2}}
\right]\,.
\end{align}
\end{widetext}
An equivalent expression can be obtained by isolating the $m=0$ contribution in the 
sum rather than the $n=0$ term. The first sum in Eq.~\eqref{eq:g0_2} can be evaluated 
easily by using:
\begin{equation}\label{eq:sum1}
\sum_{k=1}^{\infty} \frac{\cos kx}{k^{2}} = \frac{\pi^{2}}{6}-\frac{\pi|x|}{2}+\frac{x^{2}}{4}\,.
\end{equation}
Thus we have:
\begin{align}\label{eq:harmonic_poly}
H(y-\eta) 
&= -\frac{2}{p_{1}p_{2}}\sum_{m=1}^{\infty}\frac{\cos\frac{2\pi m}{p_{2}}(y-\eta)}{(\frac{2\pi m}{p_{2}})^{2}}\notag\\[7pt]
&= -\frac{1}{2\,p_{1}p_{2}}\left(\frac{p_{2}^{2}}{6}-p_{2}|y-\eta|+|y-\eta|^{2}\right)\,. 
\end{align}
The second sum in Eq. \eqref{eq:g0_2} can be evaluated with the help of the Poisson 
summation formula:
\begin{equation}\label{eq:cosine_series}
\sum_{m=-\infty}^{\infty} f(m)\cos pm = \sum_{k=-\infty}^{\infty}\int_{-\infty}^{\infty} dt\,f(t)\cos(2k\pi+p)\,t\,.
\end{equation}
In particular, if we choose:
\begin{gather*}
p = \frac{2\pi}{p_{2}}(y-\eta)\,,\qquad
f(m) = \frac{1}{\left(\frac{2\pi n}{p_{1}}\right)^{2}+\left(\frac{2\pi m}{p_{2}}\right)^{2}}\,,
\end{gather*}
we can write the second sum in Eq.~\eqref{eq:g0_2} as:
\begin{widetext}
\begin{align}\label{eq:g0_3}
K(x-\xi,y-\eta) 
&= -\frac{2}{p_{1}p_{2}}
\sum_{n=1}^{\infty}\cos\frac{2\pi n}{p_{1}}(x-\xi)
\sum_{m=-\infty}^{\infty}\cos\frac{\cos\frac{2\pi m}{p_{2}}(y-\eta)}
{\left(\frac{2\pi n}{p_{1}}\right)^{2}+\left(\frac{2\pi m}{p_{2}}\right)^{2}}\notag\\[7pt]
&= -\frac{2}{p_{1}p_{2}}
\sum_{n=1}^{\infty}\cos\frac{2\pi n}{p_{1}}(x-\xi)
\sum_{k=-\infty}^{\infty}\int_{-\infty}^{\infty}dt\,\frac{\cos(2\pi k+p)t}
{\left(\frac{2\pi n}{p_{1}}\right)^{2}+\left(\frac{2\pi t}{p_{2}}\right)^{2}}
\end{align}
\end{widetext}
The integral can be easily calculated by considering:
\[
\int_{0}^{\infty} dx\,\frac{\cos \omega x}{a^{2}+x^{2}}=\frac{\pi}{2a}\,e^{-\omega a}\,.
\]
Thus Eq.~\eqref{eq:g0_3} becomes:
\begin{multline}\label{eq:g0_4}
K(x-\xi,y-\eta) = -\frac{1}{2\pi}\sum_{k=-\infty}^{\infty}\sum_{n=1}^{\infty}
\frac{e^{-\frac{2\pi n}{p_{1}}|p_{2}k+y-\eta|}}{n}\\\cos\frac{2\pi n}{p_{1}}(x-\xi)
\end{multline}
The sum in $n$ can be calculated by noting:
\begin{equation}\label{eq:sum_log}
\sum_{n=1}^{\infty}\frac{e^{-2\pi n x}}{n}\,\cos(2\pi n y)
= -\log|1-e^{-\sigma}|\,,
\end{equation}
where $\sigma=2\pi(x\pm iy)$ with $x>0$ and arbitrary choice of sign.
Thus separating positive and negative values of $k$ in Eq.~\eqref{eq:g0_4} 
and using Eq. \eqref{eq:sum_log} the function $K(x-\xi,y-\eta)$ takes the
final form:
\begin{multline}\label{eq:g0_6}
2\pi K(x-\xi,y-\eta)
=\log\left|1-e^{\frac{2\pi i}{p_{1}}(z-\zeta)}\right|\\[7pt]
+\sum_{k=1}^{\infty}\log\left|1-2q^{2k}\cos\frac{2\pi}{p_{1}}(z-\zeta)+q^{4k}\right|\,,
\end{multline}
where:
\[
\left\{
\begin{array}{l}
z = x+iy\\
\zeta = \xi+i\eta
\end{array}
\right.\,,
\qquad
q = e^{-\frac{\pi p_{2}}{p_{1}}}\,.
\]
The second term in Eq.~\eqref{eq:g0_6} can be expressed in terms of the Jacobi theta 
function $\vartheta_{1}(u,q)$ defined as:
\[
\vartheta_{1}(u,q)
= 2q^{\frac{1}{4}}\sin u \prod_{n=1}^{\infty}\left(1-2q^{2n}\cos2u+q^{4n}\right)\left(1-q^{2n}\right)\,.
\]
Another useful relation can be obtain by taking the derivative of $\vartheta_{1}(u,q)$ 
with respect to $u$:
\begin{equation}
\lim_{u\rightarrow 0}\frac{\vartheta_{1}(u,q)}{\sin u}
=\lim_{u\rightarrow 0}\frac{\vartheta_{1}'(u,q)}{\cos u}
=\vartheta_{1}'(0,q)\,.
\end{equation}
Thus we have:
\begin{equation}
\vartheta_{1}'(0,q)
= \lim_{u\rightarrow 0}\frac{\vartheta(u,q)}{\sin u}
= 2q^{\frac{1}{4}}\prod_{k=1}^{\infty}\left(1-q^{2k}\right)^{3}\,.
\end{equation}
In this way we can express:
\begin{equation}
\prod_{k=1}^{\infty}(1-2q^{2k}\cos 2u +q^{4k})
= \frac{1}{(2q^{\frac{1}{4}})^{\frac{2}{3}}\sin u}
\left[\frac{\vartheta_{1}(u,q)}{\vartheta_{1}'^{\frac{1}{3}}(0,q)}\right]\,.
\end{equation}
Then taking: 
\[
u = \frac{\pi}{p_{1}}(z-\zeta)
\]
and substituting in Eq.~\eqref{eq:g0_6} we obtain:
\begin{multline}\label{eq:g0_7}
2\pi K(x-\xi,y-\eta)
=\frac{\log 2}{3}
+\frac{\pi}{6}\frac{p_{2}}{p_{1}}\\[7pt]
-\frac{\pi}{p_{1}}|y-\eta|
+\log\left|\frac{\vartheta_{1}(u,q)}{\vartheta_{1}'^{\frac{1}{3}}(0,q)}\right|\,.
\end{multline}
Combining Eq.~\eqref{eq:g0_7} with Eq.~\eqref{eq:harmonic_poly} we conclude
that:
\begin{align}
G_{0}(\bm{x},\bm{y})
&= H(y-\eta)+K(x-\xi,y-\eta)
\notag\\[7pt]
&= \frac{\log 2}{6\pi}
- \frac{1}{2\,p_{1}p_{2}}|y-\eta|^{2}
+ \frac{1}{2\pi}\log\left|\frac{\vartheta_{1}(u,q)}{\vartheta_{1}'^{\frac{1}{3}}(0,q)}\right|\,.
\end{align}
An alternative notation frequently used for the Jacobi theta function is:
\[
\vartheta_{1}(u|\tau)=\vartheta(u,q)\qquad q=e^{i\pi\tau}\,.
\]
With this choice we can write the Green function in the final form:
\begin{multline}\label{eq:final_green_function}
G_{0}(\bm{x},\bm{y})
= \frac{\log 2}{6\pi}
-\frac{1}{2\,p_{1}p_{2}}|y-\eta|^{2}\\[7pt]
+\frac{1}{2\pi}\log\left|\frac{\vartheta_{1}(\frac{z-\zeta}{p_{1}/\pi}|\frac{ip_{2}}{p_{1}})}
 {\vartheta_{1}'^{\frac{1}{3}}(0|\frac{ip_{2}}{p_{1}})}\right|\,.
\end{multline}

\section{\label{app:2}Derivation of the Functions $\Gamma_{s}(\bm{x})$ and $\Gamma_{d}(\bm{x},\bm{x}_{k})$}

In this appendix we derive the analytical expression for the stress
functions $\Gamma_{s}(x)$ and $\Gamma_{d}(\bm{x},\bm{x}_{k})$ in 
Eq.~\eqref{eq:gamma_total}. The former is given by the integral:
\begin{equation}\label{eq:gs_1}
\frac{\Gamma_{s}(\bm{x})}{Y} = \int d^{2}y\,K(\bm{y})[G_{0}(\bm{x},\bm{y})-\langle G_{0}(\cdot,\bm{y})\rangle]
\end{equation}
It is convenient to keep the Green function $G_{0}(\bm{x},\bm{y})$
in the form of Eq.~\eqref{eq:g0_2}. Thus we have:
\begin{align*}
I_{1}
&= \int d^{2}y\,K(\bm{y})G_{0}(\bm{x},\bm{y})\\
&=-\frac{p_{1}}{2\pi^{2}} \int_{-\pi}^{\pi}d\psi'\,\cos\psi'\sum_{n=1}^{\infty}\frac{\cos\frac{2\pi n}{p_{1}}(\xi-\xi')}{n^{2}}\,,
\end{align*}
which using Eq.~\eqref{eq:sum1} and carrying out the integrals 
yields:
\begin{equation}\label{eq:gs_2}
I_{1}=\log\left[\frac{2(r^{2}-1)}{(r+\cos\psi)(r+\sqrt{r^{2}-1})}\right]
\end{equation}
To calculate the second integral in Eq.~\eqref{eq:gs_1} it is 
convenient to invert the order of integration and use the result
of Eq.~\eqref{eq:gs_2}:
\begin{align}\label{eq:gs_3}
I_{2}
&= \int d^{2}y\,K(\bm{y})\langle G_{0}(\cdot,\bm{y})\rangle
 = \int \frac{d^{2}y}{A}\,\Gamma_{s,1}(\bm{y}) \notag\\[7pt]
&= \log\left[\frac{4(r^{2}-1)}{(r+\sqrt{r^{2}-1})^{2}}\right]-\frac{r-\sqrt{r^{2}-1}}{r}\,.
\end{align}
Combining Eq.~\eqref{eq:gs_2} and \eqref{eq:gs_3} we obtain:
\begin{equation}\label{eq:gs_4}
\frac{\Gamma_{s}(\bm{x})}{Y}
= \log\left[\frac{r+\sqrt{r^{2}-1}}{2(r+\cos\psi)}\right]
+ \frac{r-\sqrt{r^{2}-1}}{r}\,.
\end{equation}

The elastic stress produced at the point $\bm{x}$ by a disclination at 
$\bm{x}_{k}$ is given by the function:
\begin{equation}\label{eq:gd_1}
\frac{\Gamma_{d}(\bm{x},\bm{x}_{k})}{Y} 
= G_{0}(\bm{x},\bm{x}_{k})-\langle G_{0}(\cdot,\bm{x}_{k})\rangle\,.
\end{equation}
To calculated the average $\langle G_{0}(\cdot,\bm{x})\rangle$ one can 
again start from $G_{0}(\bm{x},\bm{y})$ in the form of a series and 
use Eq.~\eqref{eq:sum1}. This yields
\begin{multline}\label{eq:gd_2}
A\langle G_{0}(\cdot,\bm{x})\rangle 
= -\frac{\pi}{6}R_{1}R_{2}
  +\frac{1}{2}\int_{-\pi}^{\pi}d\psi'\sqrt{g}\,|\xi-\xi'|\\
  -\frac{1}{2p_{1}}\int_{-\pi}^{\pi}d\psi'\,\sqrt{g}\,|\xi-\xi'|^{2}\,.
\end{multline}
To calculate the integrals in Eq.~\eqref{eq:gd_2} it is convenient to
expand the conformal angle $\xi$ in Fourier harmonics:
\[
\xi(\psi) = \sum_{n=1}^{\infty} b_{n}\sin n\psi
\]
with:
\begin{equation}\label{eq:fourier1}
b_{n} 
= \frac{\kappa}{n}\,[\alpha^{n}-(-1)^{n}]
\end{equation}
where $\alpha=\sqrt{r^{2}-1}-r$. Thus we have:
\begin{gather}\label{eq:gd_3}
I_{3} 
=\int_{-\pi}^{\pi}d\psi'\,\sqrt{g}\,|\xi-\xi'|\notag\\
=2R_{1}R_{2}\psi\xi+2R_{2}^{2}\log\left(\frac{r-1}{r+\cos\psi}\right)
+2R_{1}R_{2}\int_{\psi}^{\pi}d\psi'\,\xi'\,,
\end{gather}
with
\begin{gather}
\int_{\psi}^{\pi}d\psi'\,\xi 
= \sum_{n=1}^{\infty}\frac{b_{n}}{n}\,(\cos n\psi-\cos n\pi)\notag\\
= \frac{1}{4}\kappa(\pi^{2}-\psi^{2})-\kappa\Li(-\alpha)+\kappa\Real\{\Li(\alpha e^{i\psi})\}
\end{gather}
where $\Real\{\cdot\}$ stands for the real part and
\[
\Li(z) = \sum_{n=1}^{\infty}\frac{z^{2}}{n^{2}}
\]
is the usual Euler's dilogarithm (see Ref.~\onlinecite{AbramowitzStegun} pp 1004-1005).
The second integral in Eq.~\eqref{eq:gd_2} is give by:
\begin{gather}\label{eq:gd_4}
I_{4} 
=\int_{-\pi}^{\pi}d\psi'\,\sqrt{g}\,|\xi-\xi'|^{2}\notag\\
=R_{1}R_{2}\int_{-\pi}^{\pi}d\psi'\,|\xi-\xi'|^{2}
-2\pi\kappa R_{2}^{2}\log\left[\frac{2(r^{2}+1)}{r+\sqrt{r^{2}-1}}\right]\,.
\end{gather}
To calculate the integral in Eq.~\eqref{eq:gd_4} one uses Parseval's 
identity:
\[
\frac{1}{\pi}\int_{-\pi}^{\pi} dx\,f^{2}(x) 
= \frac{a_{0}^{2}}{2}+\sum_{n=1}^{\infty}(a_{n}^{2}+b_{n}^{2})\,,
\]
where $f(x)$ an arbitrary square-integrable function on the interval $[-\pi,\pi]$
with Fourier series:
\[
f(x) = \frac{a_{0}}{2}+\sum_{n=1}^{\infty}(a_{n}\cos nx+b_{n}\sin nx)\,.
\]
Thus we have:
\begin{align}\label{eq:gd_5}
\int_{-\pi}^{\pi}d\psi'\,|\xi-\xi'|^{2}
&=2\pi\xi^{2}+\pi\sum_{n=1}^{\infty}b_{n}^{2}\notag\\
&=\pi\kappa^{2}\left[2\Li(\alpha^{2})+\frac{\pi^{2}}{6}\right]\,.
\end{align}
Using Eq.~\eqref{eq:gd_3}, \eqref{eq:gd_4} and \eqref{eq:gd_5} we conclude that:
\begin{align}\label{eq:gd_6}
\langle G_{0}(\cdot,\bm{x}) \rangle
&=-\frac{\kappa}{16\pi^{2}}\left(\psi-\frac{2}{\kappa}\xi\right)^{2}\notag\\[5pt]
&-\frac{\kappa}{8\pi}\Li(\alpha^{2})
 +\frac{\kappa}{4\pi^{2}}\Real\{\Li(\alpha e^{i\psi})\}\notag\\[5pt]
&+\frac{1}{4\pi^{2}r}\log\left[\frac{2(r^{2}-1)}{(r+\cos\psi)(r+\sqrt{r^{2}-1})}\right]\,.
\end{align}

\section{\label{app:3}Cluster Optimization via Tapping}

The Tapping Algorithm (TA) is a hybrid algorithm designed to find the optimal crystalline structure of particle systems constrained to lie on a curved surface and interacting with a long range potential of the form $U_{ij}=1/|\bm{r}_{i}-\bm{r}_{j}|^{s}$. Hybrid algorithms, such as Basin-Hopping \cite{WalesDoye:1997}  and Minima-Hopping \cite{Goedecker:2004}, have been successfully employed throughout the years to predict the crystalline structure of molecular clusters and proteins. In general they combine fast local minimizations with global moves whose goal is to release the system from the local minimum it is confined at the end of a local minimization step. 

A typical hybrid optimization routine can be summarized in the following two steps: \emph{1}) after all the independent variables have been randomly initialized, a local optimization is performed and a local minimum $x$ is determined; \emph{2}) from $x$ a new configuration $y$ is constructed by applying a  global (generally stochastic) move. The new configuration $y$ is then used as starting point for a new local optimization step. The two steps are iterated until a stopping criterion is satisfied. The goal is thus to explore the largest possible number of local minima and avoid visiting the same minimum too often. 

The crucial point in designing an effective hybrid algorithm is clearly the choice of the global move. There is no general rule to identify a successful global transformation $x \rightarrow y$ and physical intuition and prior experience are typically the only guidelines. In the case of Basin-Hopping, for instance, the global transformation consists in a Monte Carlo move in which all the particles of the system are randomly displaced in order to construct a new initial configuration from which a new trial minimum is obtained. The step is accepted with probability $\exp(-\beta\Delta V)$, where $\Delta V$ is the energy difference between the new and previous minimum and $\beta$ is an inverse temperature adjusted to obtain a $50\%$ acceptance ratio. In the case of Minima-Hopping the escape step is performed by a short Molecular Dynamics simulation by assigning the particles a fixed kinetic energy. 

The global move adopted in TA is inspired by the process of close packing of spherical objects by tapping and is motivated by the well established role of topological defects in determining the order of two-dimensional non-Euclidean crystals as well as the picture of the potential energy surface (PES) of such systems as a multi-funnel landscape. Consider a system of say spherical objects confined in a two-dimensional box with an initial disordered configuration. A common way to bias the system toward a close-packed configuration is to provide it kinetic energy by gently tapping the box. If the system is populated by locally ordered regions (i.e. grains) separated by clusters of defects, the primary effect of tapping is to produce a glide of defects inside the crystals with a subsequent rearrangement of grains.  This mechanism can be reproduced numerically in the following way. The algorithm starts with a random distribution of particles and rapidly quenches the system by performing a fast local minimization. Once particles are trapped in a local minimum, defects are identified by a Delaunay triangulation of the lattice. Then the system is tapped by adding to the defect positions a random displacement. The magnitude of the displacement is given by the typical spacing associated with the particles number times a factor $\lambda$ which represents the tapping strength. This factor is initially set to $10^{-3}$. After defects have been moved a new local minimization is performed in order to construct the trial configuration $y$. The energy of this configuration is compared with the energy of the previous minimum and the move is accepted if their difference is larger than some tolerance factor $\epsilon_{E}$. If, on the other hand, the energy difference is smaller than $\epsilon_{E}$, the system has relaxed again to the same minimum. In this case the tapping strength is increased of a factor 10 and the process is repeated until the system successfully hops to a new minimum. The tapping strength $\lambda$ is then set to its initial value. The process is iterated until the rate of discovery of new global minima drops below some threshold value or a maximum number of iterations is reached. 

In the current implementation of the algorithm, the local minimization step is performed using the Fletcher-Reeves conjugate gradient algorithm \cite{PressEtAl:1992}. Analytic expressions for the energy gradient and the Hessian matrix are coded in the program in order to reduce the number of evaluations of the objective function during the relaxation step to one single event. The Delaunay triangulation is calculated via the Dwyer's divide and conquer algorithm with alternate cuts~\cite{Dwyer:1987}, which runs in $O(N\log\log N)$ time, making the identification of the defects particularly fast.

The main difference between TA and other hybrid algorithms (including Basin-Hopping) is that the escape move consists of adaptive displacements of defects only, rather than of the entire system. In the case of non-Euclidean crystals, where the conformation of the energy landscape is subtly related to the arrangement of topological defects, this mechanism is believed to explore the PES more accurately. In systems as Lennard-Jones clusters or spin-glasses, the PES is characterized by an exponential number of local minima separated by energy barriers. For this reason the majority of the algorithms are specifically designed to allow the system to overcome a barrier by providing it a significant amount of energy. If the energy landscape, however, is characterized by the presence of multiple narrow funnels, as believed in this case, the previous methods become ineffective. A funnel represents the basin of attraction of a given local minimum. If the global minimum is also located at the bottom of a funnel, an algorithm that is attempting to locate it via a sequence of local minimization steps has a chance to find it exclusively by starting from a configuration already at the muzzle of the funnel. Such possibility, however, is ruled out if all the particles are displaced simultaneously during the escape move and the system is abruptly moved to a completely different place in the energy landscape. On the other hand, by adaptively tapping the defects it is possible to achieve a much finer inspection of the PES and possibly locate the funnel associated with the global minimum. A copy of our code is available by request.

\end{document}